\documentclass[%
twocolumn,
superscriptaddress,
amsmath,amssymb,
aps,
prl,
]{revtex4-2}

\usepackage[utf8]{inputenc}
\usepackage{CJKutf8}
\usepackage{graphicx}
\usepackage{subfigure}
\usepackage{color}
\usepackage[dvipsnames]{xcolor}
\usepackage{amsmath,amssymb,bm}
\usepackage{cases}
\usepackage{tikz}
  \usetikzlibrary{calc}
  \usetikzlibrary{arrows}
  \usetikzlibrary{decorations.markings}
\usepackage{afterpage}
\usepackage{ulem}
\usepackage{hyperref}
\usepackage{xcolor} 
\definecolor{mydarkred}{RGB}{139,0,0} 
\expandafter\ifx\csname package@font\endcsname\relax\else
  \expandafter\expandafter
  \expandafter\usepackage
  \expandafter\expandafter
  \expandafter{\csname package@font\endcsname}
\fi

\newcommand{\etal}{{\it et al.}}

\newcommand{\elabel}[1]{\label{eq:#1}}
\newcommand{\eref}[1]{Eq.~(\ref{eq:#1})}

\newcommand{\Eref}[1]{Eq.~(\ref{eq:#1})}

\usepackage{dsfont}

\newcommand{\canetset}[1]{{\mathchoice {\hbox{$\sf\textstyle #1\kern-0.4em #1$}}
{\hbox{$\sf\textstyle #1\kern-0.4em #1$}}
{\hbox{$\sf\scriptstyle #1\kern-0.3em #1$}}
{\hbox{$\sf\scriptscriptstyle #1\kern-0.2em #1$}}}}

\begin{document}

\title{Quantum Analog of Vicsek Model for Active Matter}

\author{Hong Yuan (\begin{CJK}{UTF8}{gbsn}袁红\end{CJK})}
\affiliation{Graduate School of CAEP, Beijing 100193, China}

\author{L.X. Cui (\begin{CJK}{UTF8}{gbsn}崔廉相\end{CJK})}
\affiliation{Beijing Computational Science Research Center, Beijing 100193, China}

\author{L.T. Chen (\begin{CJK}{UTF8}{gbsn}陈乐天\end{CJK})}
\affiliation{Department of Mathematics and Centre of Complexity Science, Imperial College London, South Kensington, London SW7 2BZ, United Kingdom}

\author{C.P. Sun (\begin{CJK}{UTF8}{gbsn}孙昌璞\end{CJK})}
\email{suncp@gscaep.ac.cn}
\affiliation{Graduate School of CAEP, Beijing 100193, China}

\begin{abstract}
We propose a quantum model consisting of an ensemble of overdamped spin$-1/2$ particles with ferromagnetic couplings, driven by a radially homogeneous magnetic field. The spontaneous magnetization of the spin components breaks the $SO(3)$ (or $SO(2)$) symmetry, inducing an ordered phase of flocking. Our model converges to the Vicsek model in the classical limit and corresponds to the Toner-Tu model in the continuous limit. Our investigation not only elucidates the intrinsic connection between these two models, but also introduces new opportunities for exploring the mechanisms underlying flocking order and correlations at the quantum level, which maybe pave the way for a new field of research---the quantum active matter.
\end{abstract}

\maketitle

\textit{Introduction.}---Recent intensive studies of active matter have revolutionized our understanding of non-equilibrium systems by revealing the spontaneous emergence of long-range polar order among locally aligning, self-propelled particles, especially in two dimensions \cite{Gompper_2020,Bechinger_2016,Ramaswamy_2017,viscek_model,Chate_2008},  which are similar to the flocking behavior observed in bird flocks, fish schools, and even synthetic micro-swimmers \cite{Ballerini_2008,Shaebani_2020}. In a pioneering work, Vicsek \etal \ \cite{viscek_model} introduced a simple yet powerful model demonstrating that even with minimal local rules, particles could spontaneously align to form long-range polar order. Another fundamental breakthrough was made almost simultaneously by Toner and Tu (TT) in 1995 \cite{Toner_Tu}, developing a hydrodynamic theory of active matter. In TT theory, the scaling exponents, which characterize the temporal and spatial correlations of fluctuations, could be computed exactly, contrasting with predictions by the Mermin-Wagner-Hohenberg theorem \cite{MerminWagner,Hohenberg_67}. This seminal work laid the foundation for understanding the collective dynamics of active matter \cite{John,Ginelli}.

Despite the advantages offered by the Vicsek model and TT theory, subsequent studies revealed certain limitations \cite{Chate_2024}. For example, Toner himself later realized that the original hydrodynamic analysis was incomplete due to the neglect of certain important terms, undermining most claims of exactness in the original analysis \cite{Toner2012}. There remained hope, however, that renormalization could render these additional terms irrelevant. This expectation was challenged by recent large-scale simulations of the Vicsek model \cite{Mahault2019}. These simulations exhibited scaling behavior that clearly deviated from the predictions of Toner and Tu, highlighting the need for a refined theoretical framework.

Our study aims to bridge this gap by proposing a quantum analog of the Vicsek model, named the quantum Vicsek model (QVM). We consider an ensemble of overdamped spin$-1/2$ particles with ferromagnetic couplings, driven by a radially homogeneous magnetic field. The Heisenberg interactions in this system typically exhibit a collective behavior—spontaneous magnetization, breaking the $SO(3)$ (or $SO(2)$ in $2D$) symmetry in $3D$, and inducing spatial motion into an ordered phase analogous to classical flocking. This can also be seen as a flying XY model in $2D$ \cite{Toner_Tu}.
Our model primarily operates in the quantum regime, but it can be applied to the classical case under the mean-field and high temperature approximations \cite{sun_yu_1994,sun_yu_1995}.
Specifically, by taking the continuum limit of the QVM, we derive the hydrodynamic equations for active matter using a mean-field approximation up to the next leading order. The similarity between our hydrodynamic equation from the quantum Vicsek model and the Toner-Tu equation is reflected by the detailed calculations of critical scaling exponents according to the renormalization group. Our methodology for this continuum treatment incorporates quantum aspects absent in classical models, and the quantum perspective allows for a deeper understanding of the microscopic mechanisms driving the collective behavior of active particles, explaining why the results from the Vicsek Model and the Toner-Tu hydrodynamics may belong to the same universality class at the microscopic level. Moreover, our study suggests a possible origin for self-propulsion, rooted in the internal spin degrees of freedom. This discovery opens the door to studying active matter through the lens of traditional thermodynamics and statistical physics, thereby laying the groundwork for further exploration for quantum effects in living systems \cite{Davies}.

\begin{figure}[h]
    \centering
    \includegraphics[width=0.7\linewidth]{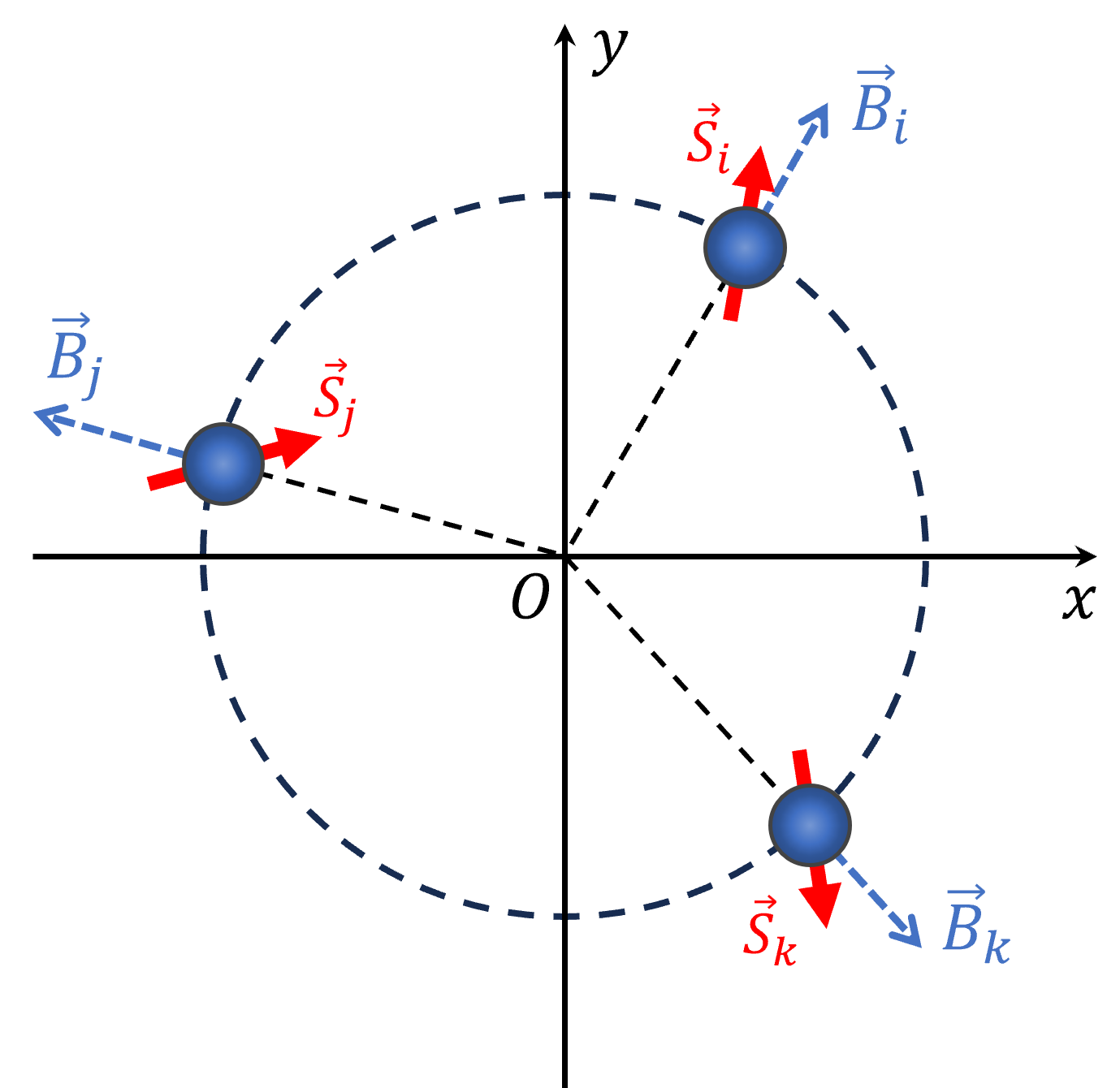}
    \caption{Spins in a radially homogeneous magnetic field.}
    \label{fig:spin}
\end{figure}

\textit{Quantum Vicsek model.}---We consider a system consisting of $N$ spin$-1/2$ particles with coordinate  $\Vec{r}_j$ and momentum $\Vec{p}_j$.
The inter-spin ferromagnetic couplings with strength $J$ has an interaction range $V_d$. There exists approximately $N' = V_d \rho$ neighboring particles interacting with spin $\vec{S}_j$, where $\rho$ is the particle density. Moreover, the entire system is subjected to a radially homogeneous magnetic $\Vec{B} \sim \Vec{r}_j$ (see Fig.~\ref{fig:spin}). The Hamiltonian of the system reads
\begin{equation}
\elabel{Hamiltonian}
H = \sum_{j=1}^{N} \frac{p_j^2}{2m}\ -\zeta \sum_{j=1}^{N} \Vec{r_j}\cdot\Vec{S_j}\ -J\sum_{i<j}^{N'} \Vec{S_i}\cdot\Vec{S_j},
\end{equation}
where $m$ is the mass of each particle, and $\zeta$ a constant characterizing the strength of the interaction between the spins and the magnetic field. For a two-dimensional system, a radially homogeneous field can be effectively realized by placing two identical, co-axial Helmholtz coils with their magnetic moments oriented in opposite directions. The magnetic field on the midplane (perpendicular to the common axis) between the coils can be shown to exhibit local radial homogeneity. In three dimensions, a spatially dependent magnetic field can be locally approximated by its first-order Taylor expansion in position, yielding an effectively local radially homogeneous field within a small spatial region. We emphasize, however, that the main focus of this work is
the theoretical construction of a minimal microscopic model that bridges the Vicsek model and the Toner-Tu theory.
It's important to note that, as presented in the Hamiltonian, \Eref{Hamiltonian}, the interactions in this model are fully quantum mechanical and satisfy the standard commutation relations—both the spin-spin coupling and the spin-orbit coupling—similar to those in the transverse-field Ising model and other quantum spin systems. These commutation relations are essential: without them, the quantum Langevin equations presented below, which govern the dynamics of our system, could not be derived.

We assume an environment with temperature $\beta^{-1}$ coupled with this system of spin$-1/2$ particles. The dissipate dynamics \cite{Caldeira_Leggett, Leggett_RevModPhys, sun_yu_1995,sun_yu_1994,Weiss_Quantum_dissipative_systems,Breuer_Petruccione} of the spin and spatial degrees of freedom follow
\begin{align}
\elabel{langevin_equation1}
& \ddot{\vec{r}}_j(t) = \zeta{\vec{S}}_j - \gamma \dot{\vec{r}}_j + \hat{\vec{F}}_{j}(t),\\
\elabel{langevin_equation2}
& \dot{\vec{S}}_j(t) = -\left(J \sum_{i=1}^{N'} \vec{S}_i + \zeta \vec{r}_j\right) \times \vec{S}_j - \Gamma_S \cdot (\vec{S}_j - \vec{S}_j^{\mathrm{st}}) \notag \\
& \hspace{3.5em} + \hat{\vec{F}}_{Sj}(t),
\end{align}
where $\gamma$ is the friction coefficient for the spatial part, the dissipation tensor $\Gamma_S$ of the spin part is $\Gamma_s=diag(2\gamma_s,2\gamma_s,\gamma_s)$ where $z-$axis is along $\Vec{S}_j^{st}$, the stationary value of $\Vec{S}_j$. We emphasize that $\gamma_s$ is the friction coefficint for the spin part, and in general relate to the temperature of environment. 
Both Brownian forces $\hat{\vec{F}}_{j}(t)$ and $\hat{\vec{F}}_{Sj}(t)$ need to adhere to the fluctuation-dissipation theorem (FDT), i.e., $\langle \{\hat{F}_{ja}(t), \hat{F}_{ib}(t')\}\rangle_R = \delta_{ij}\delta_{ab}\Phi(t-t')$ and $\langle \hat{F}^a_{Sj}(t) \hat{F}^b_{Si}(t) \rangle_R = \hbar^2\gamma_s\delta_{ij}\delta_{ab}\delta(t-t')/8$ (see Supplemental Material, Sec. SI \cite{SM_This} for the explicit form of $\Phi(t)$). Here, \(a, b\) are Cartesian indices for dimensions --- \(x, y\) in $2D$ (or \(x, y, z\) in $3D$), $\{\cdot\}$ represents the anticommutator, and \(\langle \cdot \rangle_R\) denotes averaging over the environment. The detailed derivation of the quantum Langevin equations \Eref{langevin_equation1} and \Eref{langevin_equation2} is presented in Supplemental Material, Sec. SI \cite{SM_This}.

With stronger inter-spin coupling of ferromagnetic type, i.e., a large positive $J$, the spontaneous breaking of $SO(3)$ symmetry emerges to give a mean field
\begin{equation}
\elabel{mean_field_B}
 \Vec{B}_{MF}=\frac{1}{2}J\sum_{j}^{N'}\langle \Vec{S}_j\rangle\equiv \frac{1}{2}JN'\Vec{S},
\end{equation}
where $\langle\cdot\rangle$ means some kind of average over classical or quantum states of the system \cite{haken1986laser}.
In the overdamped regime, the inertial terms can be neglected. Under the mean-field approximation where the commutation relations effectively vanish, and in the high temperature limit, \Eref{langevin_equation1} and \Eref{langevin_equation2} reduce to their classical forms \cite{haken1986laser}
 \begin{align}
\elabel{overdamped_EOM1}
&\dot{\overrightarrow{r_j}} = u \vec{S}_j + \vec{f}(t),\\
\elabel{overdamped_EOM2}
&  \dot{\vec{S}}_j = -\vec{B}_{MF}\times \vec{S}_j-\Gamma_S \cdot (\vec{S}_j-\vec{S}),
 \end{align}
where $\langle f_{ja}(t) f_{ib}(t')\rangle_R = 2/(\gamma m\beta)\delta_{ij}\delta_{ab}\delta(t-t')$. Here $u\equiv\zeta /m$. The spatial term $\zeta\Vec{r}_j$ is regarded as infinitesimal compared to $\Vec{B}_{MF}$ in the thermodynamic limit. As illustrated in \Eref{overdamped_EOM1}, the spin $\vec{S}_j$ acts as an energy pump for the motion of the center of mass, resulting in a nonzero mean velocity aligned with $\vec{S}_j$ due to spin-orbit coupling induced by the magnetic field  $\Vec{B}$. Moreover, ferromagnetic interactions typically align $\vec{S}_j$ with surrounding spins, leading to the emergence of an ordered phase---spontaneous magnetization---that breaks the $SO(3)$ symmetry (or $SO(2)$ in $2D$). As a result, nearly all the spins may point in the same direction $\Vec{S}$. In the overdamped regime, this ferromagnetic coupling also induces an alignment interaction among velocities, whereby a particle reorients itself according to the average direction of its neighbors within the interaction range $V_d$. This mechanism could therefore result in an ordered flocking state for spatial motions of the ensemble, propelling the entire system towards a specific direction paralleling to $\Vec{B}_{MF}$.

\begin{figure}[h]
    \centering
    \includegraphics[width=1.0\linewidth]{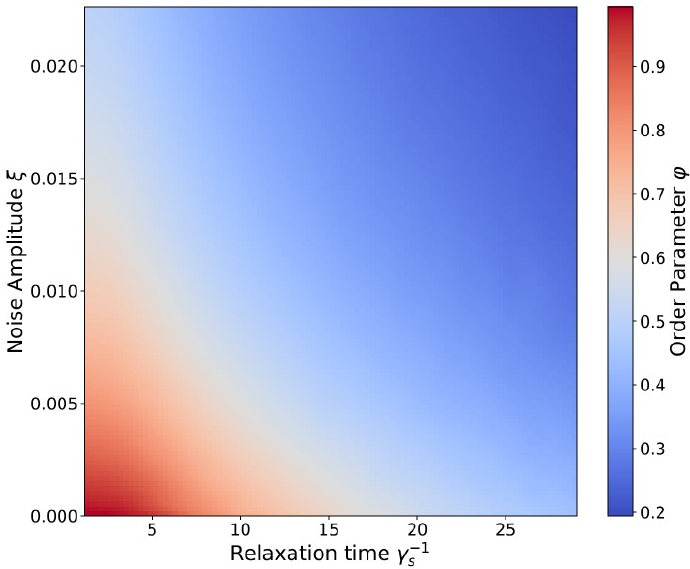}
    \caption{Numerical simulations of Quantum Vicsek Model (QVM). The phase diagram of the order parameter $\varphi=|\langle \dot{\Vec{r}}_j \rangle _j|$ as a function of the relaxation time $\gamma_s^{-1}$ and noise amplitude $\xi=\frac{2}{\gamma m \beta}$. The average of the order parameter is taking over $10^4$ time steps, and the time interval for each step is $( \Delta t = 0.1 )$. Simulation parameters: $\rho = 0.5$, $u = 0.5$, $r_c = 1.0$. }
    \label{fig:numerical_simulations}
\end{figure}

It is noteworthy that the dynamic equations \Eref{overdamped_EOM1} and \Eref{overdamped_EOM2} are consistent with the dynamics of the Vicsek model \cite{viscek_model,Ginelli}, where $u \Vec{S}_j$ acts as the self-propulsion term. In this framework, each agent moves with a constant speed $u$, and its direction $\Vec{S}_j$ tends to align with neighboring particles within a certain interaction range, as shown in \Eref{overdamped_EOM2}. Noise is present in both the magnitude and direction of the velocity.
Our model explicitly clarifies the microscopic origins of both the self-propulsion and the alignment interactions. Thus, the quantum framework presented here provides a more fundamental understanding of the Vicsek model, offering a deep insight into the studies of active matter from a microscopic perspective. We note that while earlier studies have introduced spin-like variables to describe active matter in classical contexts \cite{cavagna2015} and have explored quantum active systems using non-Hermitian models \cite{Takasan2024,Yamagishi2024}, our approach differs in several key aspects. In \cite{cavagna2015}, the so-called ``spin” variable does not represent a quantum spin, but rather serves as a classical angular momentum, and the alignment interaction is introduced phenomenologically. In contrast, our model employs quantum spin degrees of freedom, with interactions—both the Heisenberg and spin-orbit couplings—that are well-established in quantum physics. Compared to non-Hermitian quantum models \cite{Takasan2024,Yamagishi2024}, our construction is fully Hermitian, see \Eref{Hamiltonian}, ensuring a clear microscopic foundation and allowing the activity and alignment to emerge naturally from the underlying physics, without ad hoc assumptions. This Hermitian property ensures that our model can be rigorously treated within the framework of quantum mechanics.

To provide an intuitive demonstration of the QVM, we perform numerical simulations of the classical equations \Eref{overdamped_EOM1} and \Eref{overdamped_EOM2} in $3D$ space, which are derived from the quantum equations \Eref{langevin_equation1} and \Eref{langevin_equation2} under a mean-field approximation and in the high temperature limit.
The mean-field approximation is applied to particles within a radius $r_c$ around each particle. Fixing the particle number density $\rho$, the model contains two parameters: fluctuations  $\Vec{f}$ and dissipation terms $\Gamma_s$ of spin. Spin dissipation is generally characterized by the relaxation time which is characterized by $\gamma_s^{-1}$, while fluctuations are modeled as white noise with amplitude $\xi=\frac{2}{\gamma m \beta}$. 
As illustrated in Fig.~\ref{fig:numerical_simulations}, as the relaxation time decreases and the noise intensity is reduced, the order parameter $\varphi=|\langle \dot{\Vec{r}}_j \rangle_j|$, where $\langle \cdot \rangle_j$ represents the average over the whole population, transitions from a disordered phase to an ordered phase. This demonstrates the same collective behavior as in the Vicsek model.

\textit{Hydrodynamic descriptions.}---In order to understand how the Vicsek model relates to the TT theory at the microscopic level, we further consider the hydrodynamic descriptions from the QVM. To this end, we derive the hydrodynamic equation through the kinetic approach \cite{de2013non,balescu1997statistical,PhysRevE.91.033303,tokatly2000hydrodynamics} to the QVM.

Under the mean-field approximation, this interacting system is usually described by a single-body distribution \(f(\Vec{r}, \Vec{v}, t)\), which defines the particle number density in single-particle phase space at time \(t\) \cite{de2013non,balescu1997statistical}. The expected hydrodynamic variables \cite{de2013non,balescu1997statistical,PhysRevE.91.033303,tokatly2000hydrodynamics}, i.e., the spatial density and the velocity field, are then respectively given by \(\rho(\Vec{r}, t) = \int f(\Vec{r}, \Vec{v}, t) d^3\Vec{v}\) and \(\Vec{V}(\Vec{r}, t) = \int \Vec{v}f(\Vec{r}, \Vec{v}, t) d^3\Vec{v}\), where \(f(\Vec{r}, \Vec{v}, t)\) satisfies the Boltzmann equation 
\begin{equation}
\elabel{Boltzmann_equation}
\partial_t f +\Vec{v}\cdot\nabla_r f + \Vec{a}\cdot\nabla_v f =I[f],
\end{equation}
where $\partial_t$, \(\nabla_r\) and \(\nabla_v\) respectively are the differential operators with respect to time $t$, \(\Vec{r}\) and \(\Vec{v}\). \(I[f]\), following the Fokker-Planck theory \cite{van1992stochastic,risken1996fokker}, is given by $I[f]=\gamma\nabla_v\cdot(\Vec{v}-V(\Vec{r},t)+2\nabla_v/m\beta)f$. The term \(m\Vec{a}\) is the mean force acting on each particle.

In the overdamped regime, we could approximately replace $\Vec{S}_j$ with $\Vec{v}_j$ in Hamiltonian \Eref{Hamiltonian}. The accuracy of this approximation is enhanced when the symmetry $SO(3)$ spontaneously breaks in the spin part, leading to accordingly phase transition in velocities, with $\Vec{V}(\Vec{r},t)$ being the order parameter. In the mean-field approximation, the energy reads
\begin{equation}
\elabel{energy}
H_{ov}=E_0[N]+\sum_{j=1}^{N}[\frac{mv_j^2}{2}-\frac{J}{2u^2}N'\Vec{V}\cdot\Vec{v}_j],
\end{equation}
with the constraint $|\Vec{v}_j|=u$. Here the last term results from the Heisenberg interaction, and $E_0[N]$ only depends on the particle numbers. To apply  \Eref{Boltzmann_equation} to QVM, we need to determine the free energy by calculating the partition function while taking into account that nearly all the \(\Vec{v}_j\)s are aligned with \(\Vec{V}\). In the high temperature limit, namely, when the average velocity of particles is much smaller than that of the thermal motion, we expand the free energy density to the second order of $\beta mV^2$ as
\begin{equation}
\elabel{free_energy}
F(\Vec{r}) =\rho( -\lambda|\Vec{V}|^2 +\eta|\Vec{V}|^4+F_0),
\end{equation}
where $\lambda=3J^2N'^2/8\Tilde{m}u^4$, $\eta=2\lambda^2\beta/9$ and $F_0=(3\ln{\Tilde{m}\beta}-\ln{2\pi}+2\ln\rho+2\kappa u^2)/2\beta$. Here, $\kappa$ is the Lagrange multiplier resulting from the constraint on velocities, and the renormalized mass is defined as $\Tilde{m} \equiv m + 2\kappa/\beta$. For such $\phi^4$ field, the symmetry spontaneously breaks at $|\Vec{V}_0|^2=9/(4\lambda\beta)$, which is just the order parameter \cite{landau2013statistical}. Consequently, the mean force acting on each particle is given by $m\Vec{a}=-\nabla_r(F/\rho)$.

We then obtain the evolution of $\Vec{V}(\Vec{r}, t)$  straightforwardly \cite{de2013non, balescu1997statistical, PhysRevE.91.033303,tokatly2000hydrodynamics}, namely,
\begin{align}
\elabel{hydrodynamic_equation2}
m\left(\partial_t +\lambda_1 \Vec{V} \cdot \nabla_r \right) \Vec{V} = \Tilde{\lambda} \Vec{V}-\Tilde{\eta}\lvert \Vec{V}\rvert^2\Vec{V} - \frac{1}{ \rho} \nabla_r \cdot \Hat{P}+\Vec{R}_v.
\end{align}
Here, we have considered that, when $t$ is large, the directions of velocity and position are approximately the same, i.e., $\Vec{V}=\xi \Vec{r}$.   Here, $\Tilde{\lambda}\equiv 2\xi \lambda$, $\lambda_1\equiv 1-2\lambda$ and $\Tilde{\eta}=4\xi \eta$. $\Hat{P}_{ab} = (\rho/\beta )\delta_{ab} -\rho (\gamma\beta)^{-1}(\partial_{ra} V_b+ \partial_{rb} V_a)$ are the stress tensor; $\partial_{ra}\equiv\partial/\partial r_a$ with $a,b$ are Cartesian indices, $\Vec{R}_v(\Vec{r},t)$ is a random force subjected to the FDT, i.e., $\langle R_{va}(\Vec{r},t) R_{vb}(\Vec{r}',t') \rangle_R = 2\gamma m\rho^{-1}[V_aV_b+\Hat{I}_{ab}(m\beta)^{-1}]\delta_{r_c} (\Vec{r}-\Vec{r}')\delta (t-t')$. Here $\delta_{r_c} (\Vec{r}-\Vec{r}')$ is a Dirac-$\delta$ like function implying the existence of the cutoff in the spatial freedom. The detailed discussion about the derivation of hydrodynamic equations is illustrated in Supplemental Material, Sec. SII \cite{SM_This}.  A complete hydrodynamic framework is thus formed by \Eref{hydrodynamic_equation2} together with the evolution of $\rho$, which follows the conservation law of particle number, i.e., $\partial_t\rho=-\nabla_r\cdot(\rho\Vec{V})$.
 
\textit{Flocking phase transition.}---In order to analyze the symmetry broken phase, we shift to a comoving coordinate frame moving with $\Vec{V}_0$, $\Vec{V}'=(\Vec{V}_\perp,V_\parallel \hat{e}_\parallel)$ with $\hat{e}_\parallel=\Vec{V}_0/\lvert \Vec{V}_0\rvert$. Here, $\Vec{V}_\perp(\Vec{r},t)$ is the Goldstone mode dominating the critical behavior of this system flocking. Its hydrodynamic equation then becomes

\begin{align}
\elabel{hydrodynamic_equation3}
&\left(\frac{\partial}{\partial t} + \lambda_1 \Vec{V}_\perp \cdot \Vec{\partial}_\perp \right) \Vec{V}_\perp \notag \\
&= -B \Vec{\partial}_\perp \delta \rho + D \Vec{\partial}_\perp \Vec{\partial}_\perp \cdot \Vec{V}_\perp  + D \nabla_r^2 \Vec{V}_\perp + \Vec{R}_\perp,
\end{align}
where $\partial_t\delta\rho=-A_I\Vec{\partial}_\perp\cdot\Vec{V}_\perp$, $\Vec{\partial}_\perp$ denoting the directional derivative along $\Vec{V}_\perp$, $\delta\rho=\rho-\rho_0$ with $\rho_0$ being the mean of the local number density and $B\equiv(\rho_0\beta)^{-1}$. The diffusion coefficient $D\equiv(\gamma m\beta)^{-1}$ is also subjected to the FDT. Notably, $\Vec{R}_\perp$ is the transformed random force with the covariance $\langle R_{\perp a}(\Vec{r},t) R_{\perp b}(\Vec{r}',t') \rangle_R = 2\Delta \delta_{ab}\delta_{r_c} (\Vec{r}-\Vec{r}')\delta_{\tau_c} (t-t')$ with the strength $\Delta\equiv \gamma(\rho \beta)^{-1}$. It is also emphasized that $A_I$ is the renormalized coefficient including the impact of the higher order (as well as nonlinear) term  $\Vec{\partial}_\perp(\delta\rho\Vec{V}_\perp)$. This simplification actually decouples the different wavelength components of $\delta\rho$ and $\Vec{V}_\perp$.
To find the critical scaling exponents, we next implement the dynamic renormalization group (DRG) method \cite{forster1977large,ma1975critical,ma2018modern} to \eref{hydrodynamic_equation3} (See Supplemental Material, Sec. SIII for detailed discussion \cite{SM_This}). The DRG procedure consists of two steps, i.e., eliminate the short wavelength components of $\Vec{V}_\perp$ and then apply scale transformation. Thanks to the simplification mentioned above, \Eref{hydrodynamic_equation3} becomes
\begin{widetext}
\begin{align}
\elabel{hydrodynamic_equation5}
[-i\omega +D(k^2+\Vec{k}_\perp\Vec{k}_\perp)+i\frac{A_IB}{\omega}\Vec{k}_\perp\Vec{k}_\perp]\cdot\Vec{V}(\Vec{k},\omega)=\Vec{R}(\Vec{k},\omega)-i\lambda_1\int_{q,\Omega}\Vec{V}(\Vec{k}-\Vec{q},\omega-\Omega)\cdot\Vec{q}_\perp\Vec{V}(\Vec{q},\Omega),
\end{align}
\end{widetext}
 where $\int_{q,\Omega}\equiv(2\pi)^{-4}\int d^3\Vec{q}\int_{-\infty}^{+\infty}d\Omega$. Here $\Vec{V}(\Vec{k},\omega)$ and $\Vec{R}(\Vec{k},\omega)$ being the Fourier transformation of $\Vec{V}_\perp(\Vec{r},t)$ and $\Vec{R}_\perp(\Vec{r},t)$. The corresponding covariance of $\Vec{R}(\Vec{k},\omega)$ is $\langle \Vec{R}(\Vec{k},\omega) \Vec{R}(\Vec{k}',\omega') \rangle_R = (2\pi)^4\Delta\delta(\omega-\omega')\delta(\Vec{k}-\Vec{k}')$ with $\Delta=\Delta(1-e^{-k^2/\Lambda^2})$. Here, $\Lambda$ is the ultraviolet cutoff of $k$ for the above Dirac-$\delta$ like function. 
In general, the nonlinear term $\lambda_1 \Vec{V}_\perp \cdot \Vec{\partial}_\perp \Vec{V}_\perp $ redefines the parameters of the reduced hydrodynamic equations containing only the long wavelength part. As a result, keeping up to $\mathcal{O}(\lambda_1^2)$, the nonlinear coupling strength $\lambda_1$ remains unchanged, $\lambda_I=\lambda_1$, and $A_IB$, as a whole, does not have correction either. In fact, one can conclude that $B$, which measures the response of velocity field to changes in density, remains unchanged in the new reduced hydrodynamic equation, i.e., $B_I=B$. Besides, the diffusion coefficient $D$ and the noise strength $\Delta$ are renormalized in the same way, as required by FDT, namely, $D_I=D[1+\Bar{\lambda}_1^2F(s;\Lambda,A_IB/D)]$ and $\Delta_I=\Delta[1+\Bar{\lambda}_1^2F(s;\Lambda,A_IB/D)]$
with $\Bar{\lambda}_1^2\equiv\lambda_1^2\Delta D^{-3}$. Here $F(s;\Lambda,A_IB/D)$ is a polynomial function of $s$, which, apart from a constant, contains only odd powers terms, with the lowest order being $s^1$. $s$ is a rescaling factor of space required in step 2. Namely, we rescale lengths, time, $\Vec{V}_\perp$ and $\delta\rho$ according to $\Vec{r}\rightarrow s\Vec{r}$, $t\rightarrow s^z t$, $\Vec{V}_\perp\rightarrow s^\chi\Vec{V}_\perp$ and $\delta\rho\rightarrow s^\chi_\rho\delta\rho$ respectively.

To further determine the scaling exponents $z$ and $\chi$, we then take an infinitesimal transformation, i.e., $s=1+l$ ($l\rightarrow0$), which leads to differential recursion relations for the parameters
\begin{equation}
\begin{aligned}
&dD(l)/dl=D[z-2+F_{31}\Bar{\lambda}_1^2],\\
& d\Delta(l)/dl=\Delta[z-2\chi-3+F_{31}\Bar{\lambda}_1^2],\\
& d\lambda_1/dl=\lambda_1[z+\chi-1],\\
& dB(l)/dl=B[\chi_\rho-1+z-\chi].
\end{aligned}
\label{flow_D}
\end{equation}
The scaling factors $z$, $\chi$ and $\chi_\rho$ are chosen to keep $D$, $\Delta$, $\lambda_1$ and $B$ fixed at their initial value, so that there exist the steady solutions $z=3/2$, $\chi=-1/2$ and $\chi_\rho=-1$. The negative $\chi$ just implies a true long-range order exists. Therefore, as an illustration, we successfully show that our QVM will emerge flocking, i.e., a stable order phase, in three dimensions and obtain the corresponding critical exponents through the hydrodynamic method. Apparently, the hydrodynamic equation derived from the QVM, as specified in \Eref{hydrodynamic_equation2}, is much similar to the $3D$ TT equation, while our model provides explicit formulations for the coefficients.
Therefore, it is convinced that the TT theory indeed can be regarded as the continuation of the Vicsek model, namely, as they may have the same quantum origin.

\textit{Conclusions.}---In conclusion, we have proposed a quantum analog of the Vicsek model (QVM), offering a novel perspective on the study of active matter. The original Vicsek model is recovered in the classical limit of the QVM, and the Toner-Tu-like equations, i.e., hydrodynamic equations, naturally emerge from its continuous formulation. This study provides insight into why the Toner-Tu equation may belong to the same universality class as the Vicsek model. Additionally, we have offered a more microscopic description of emergent flocking phenomena at the quantum level and suggest that self-propulsion could originate from internal spin degrees of freedom. This makes it feasible to study active matter using traditional thermodynamics and statistical physics, laying the groundwork for further exploration in this growing field.

By directly examining the critical behaviors of this quantum model in \Eref{Hamiltonian}, it becomes straightforward to analyze theoretical exponents in the $2-$dimensional case, which merits further exploration. This study represents a significant step towards unifying classical and quantum theories of active matter and opens new avenues for exploring the underlying mechanisms of flocking order and correlations at the quantum level, i.e., the quantum active matter. 
Moreover, our work sheds light on how concepts and methodologies from active matter could inform the study of other complex physical systems. For instance, a multi-electron condensed matter system, when focusing solely on spatial degrees of freedom, might be regarded as an active matter system. This perspective may pave the way for new approaches to understanding collective phenomena in diverse quantum systems.

\textit{Acknowledgment.}---The authors thank Ruoxun Zhai for useful discussions. This study is supported by the National Natural Science Foundation of China (NSFC) (Grant No. 12088101) and NSAF (Grant No. U2230402).

\bibliography{QVM}
\clearpage

\appendix






\end{document}


\title{Supplemental Material: Quantum Analog of Vicsek Model for Active Matter}

\author{Hong Yuan (\begin{CJK}{UTF8}{gbsn}袁红\end{CJK})}
\affiliation{Graduate School of CAEP, Beijing 100193, China}

\author{L.X. Cui (\begin{CJK}{UTF8}{gbsn}崔廉相\end{CJK})}
\affiliation{Beijing Computational Science Research Center, Beijing 100193, China}

\author{L.T. Chen (\begin{CJK}{UTF8}{gbsn}陈乐天\end{CJK})}
\affiliation{Department of Mathematics and Centre of Complexity Science, Imperial College London, South Kensington, London SW7 2BZ, United Kingdom}

\author{C.P. Sun (\begin{CJK}{UTF8}{gbsn}孙昌璞\end{CJK})}
\email{suncp@gscaep.ac.cn}
\affiliation{Graduate School of CAEP, Beijing 100193, China}

\begin{abstract}
This document is devoted to providing detailed derivations and supporting discussions of the main content of the letter. In Sec. SI, we provide the detailed derivation of the Langevin equations for the quantum Vicsek model (QVM). In Sec. SII, we derive the hydrodynamic equations corresponding to QVM in the overdamped regime. In Sec. SIII, we analyze the critical behavior of these equations by dynamic renormalization group method and calculate related critical exponents.
\end{abstract}

\maketitle

\section{SI. Derivation of Langevin equations}
\alabel{Derivation_of_Langevin_equations}

In this section, we present the detailed derivation of the Langevin equations (Eq. $(2)$ and Eq. $(3)$ in the main text) for the quantum Vicsek model. We consider that the spin and spatial degrees of freedom for each particle dissipate independently.
\subsection*{1. Spin Dynamics}

Consider a system interacting with an environment, with the total Hamiltonian
\begin{equation}
H = H_S + H_B + H_I,
\end{equation}
where $H_S$ is the system Hamiltonian, $H_B$ the bath, and $H_I$ the system-bath interaction, which is assumed to be weak compared to $H_S$ and $H_B$.

For an operator $X$ in the system Hilbert space, its evolution is governed by the Heisenberg equation of motion:
\begin{equation}
\dot{X} = -i [X, H] \equiv i \mathcal{L} X.
\end{equation}
We decompose the Liouvillian superoperator as
\begin{equation}
\mathcal{L} = \mathcal{L}_S + \mathcal{L}_B + \mathcal{L}_I,
\end{equation}
where
\begin{equation}
\mathcal{L}_S X \equiv -[X, H_S],
\end{equation}
and similarly for $\mathcal{L}_B$ and $\mathcal{L}_I$. The formal solution of the operator $X$ is given by
\begin{equation}
X(t) = e^{i t \mathcal{L}} X, \quad \text{with}\quad X\equiv X_0.
\end{equation}


We define an inner product for arbitrary operators $A,B \in L^2(\mathcal{H})$, where 
\begin{equation}
\mathcal{H} = \mathcal{H}_S \otimes \mathcal{H}_B
\end{equation}
is the direct product of the system and bath Hilbert spaces, as
\begin{equation}
(A,B) = \operatorname{Tr}\left(A^\dagger B \rho\right) = \langle A^\dagger B\rangle,
\end{equation}
with $\rho=\frac{1}{Z} e^{-\beta H}$. Under this definition, the superoperator $\mathcal{L}$ is Hermitian:
\begin{align}
(A, \mathcal{L}B) 
&= \operatorname{Tr}(A^\dagger \mathcal{L} B \rho) 
= -\operatorname{Tr}(A^\dagger [B,H]\rho) \nonumber \\
&= -\operatorname{Tr}([\mathcal{L}A^\dagger] B \rho) 
= (\mathcal{L}A,B).
\end{align}


For the operator space $L^2(\mathcal{H}_S)$ of the system, we consider a complete basis of Hermitian operators $\{X_1, X_2, \ldots\}$, where $X_{\mu}^\dagger = X_{\mu}$. The matrix $S$ is defined as
\begin{equation}
S_{\mu \nu} = (X_{\mu}, X_\nu) = \langle X_{\mu} X_\nu \rangle.
\end{equation}
The basis $\{X_{\mu}\}$ can be chosen such that $S$ is Hermitian,
\begin{equation}
S_{\mu \nu}^{*} = S_{\nu \mu}.
\end{equation}

We can then define the projection operator $\mathcal{P}$ onto the system operator subspace $L^2(\mathcal{H}_S)$ as follows. For any operator $O \in L^2(\mathcal{H}_S\otimes\mathcal{H}_B)$, its projection onto $L^2(\mathcal{H}_S)$ is given by
\begin{equation}
\mathcal{P}O = \sum_{\mu, \nu} (X_{\mu}, O)(S^{-1})_{\nu \mu} X_\nu.
\end{equation}

It can be shown as follows that $\mathcal{P}$ is self-adjoint, i.e., $\tilde{\mathcal{P}} = \mathcal{P}$. For any Hermitian operators $A$ and $B$, we have
\begin{align}
(\tilde{\mathcal{P}} A, B) &= \operatorname{Tr}\left[(\tilde{\mathcal{P}} A)^\dagger B \rho \right] \notag\\
&= \sum_{\mu,\nu} (X_\mu, A)^* (S^{-1})_{\mu\nu} \operatorname{Tr}(X_\nu B \rho ) \notag\\
&= \sum_{\mu,\nu} (A, X_\mu)(S^{-1})_{\mu\nu}(X_\nu, B) \notag\\
&= (A, \mathcal{P}B).
\end{align}


We now consider the dissipation of the spin degrees of freedom. We assume that the spatial and spin degrees of freedom of each particle are each coupled to independent thermal baths (assumed to be at the same temperature), such that they dissipate independently. Therefore, we can treat the spatial coordinates $\vec{r}_j$ as a parameter. The system Hilbert space $\mathcal{H}_S$ thus consists of the $N$-particle spin space, and the system Hamiltonian is given by Eq.(1) in the main text
\begin{equation}
H_S = \sum_{j=1}^{N} \frac{p_j^2}{2m}\ -\zeta \sum_{j=1}^{N} \vec{r_j}\cdot\vec{S_j}\ -J\sum_{i<j}^{N'} \vec{S_i}\cdot\vec{S_j}.
\end{equation}
Here, the basis $\{X_\mu\}$ is chosen as $\{\sigma_\mu^{(\alpha)} \mid \alpha = 1, \ldots, N;\, \mu = 1,2,3,4\}$, where $\sigma_4 = I_2$ is the $2\times 2$ identity matrix. In this basis, the matrix elements are $S_{ij}^{\mu\nu} = \langle \sigma_i^\mu \sigma_j^\nu \rangle$.

Let us consider the steady-state in which the $j$-th particle is aligned along the $z$-direction, i.e., $\langle \sigma_j^z \rangle \equiv \sigma_j^{\mathrm{st}} \to 1$, and $\langle \sigma_j^x \rangle \approx 0$, $\langle \sigma_j^y \rangle \approx 0$. $ \langle \vec{\sigma}_i \cdot \vec{\sigma}_j \rangle \propto e^{-\beta\sigma^{z}_i\sigma^{z}_j}<\sigma_j^{\mathrm{st}}$ is neglected in the following. Under these approximations, the matrix $S$ is block-diagonal with $N$ blocks,
\begin{equation}
S = \operatorname{diag}(S_1, S_2, \ldots, S_N),
\end{equation}
where the $j$-th block is given by
\begin{equation}
S_j = \begin{pmatrix}
1 & i\sigma_j^{\mathrm{st}} & 0 & 0 \\
-i\sigma_j^{\mathrm{st}} & 1 & 0 & 0 \\
0 & 0 & 1 & \sigma_j^{\mathrm{st}} \\
0 & 0 & \sigma_j^{\mathrm{st}} & 1
\end{pmatrix}.
\end{equation}
Accordingly, the inverse matrix takes the form
\begin{equation}
(S^{-1})_{ij}^{\mu\nu} = \delta_{ij} (S_j^{-1})^{\mu\nu}, \qquad
S_j^{-1} \approx
\begin{pmatrix}
1 & -i\sigma_j^{\mathrm{st}} & 0 & 0 \\
i\sigma_j^{\mathrm{st}} & 1 & 0 & 0 \\
0 & 0 & 1 & -\sigma_j^{\mathrm{st}} \\
0 & 0 & -\sigma_j^{\mathrm{st}} & 1
\end{pmatrix}.
\end{equation}


According to the above definition of the projection operator $\mathcal{P}$, we have
\begin{equation}
\mathcal{P}O = \sum_{i,j} \sum_{\mu, \nu} (\sigma_i^\mu, O)\,(S^{-1})_{ji}^{\nu\mu}\,\sigma_j^\nu = \sum_j \sum_{\mu, \nu} (\sigma_j^\mu, O)\,(S_j^{-1})^{\nu\mu}\,\sigma_j^\nu.
\end{equation}
Now, consider the evolution of the spin operator of the $j$-th particle in the presence of the environment. The equation of motion can be written as
\begin{align}
\label{spin_in_bath}
\frac{d}{dt} \sigma_j^\alpha(t) &= i \mathcal{L}\, e^{it\mathcal{L}} \sigma_j^\alpha \notag \\
&= e^{it\mathcal{L}}\, i \mathcal{P} \mathcal{L} \sigma_j^\alpha \notag \\
&\quad + \int_0^t d\tau\, e^{i(t-\tau)\mathcal{L}}\, i \mathcal{P} \mathcal{L} \, e^{i\tau(1-\mathcal{P})\mathcal{L}}\, i(1-\mathcal{P})\mathcal{L} \sigma_j^\alpha \notag \\
&\quad + e^{it(1-\mathcal{P})\mathcal{L}}\, i(1-\mathcal{P})\mathcal{L} \sigma_j^\alpha,
\end{align}
where $\alpha = 1,2,3$.

The first term on the right-hand side of Eq.~(\ref{spin_in_bath}) can be simplified as follows:
\begin{align}
e^{it\mathcal{L}}\, i\mathcal{P} \mathcal{L} \sigma_j^\alpha &= e^{it\mathcal{L}}\, i\mathcal{P} \mathcal{L} e^{-it\mathcal{L}}\, \sigma_j^\alpha(t) \notag\\
&\approx i\mathcal{P} \mathcal{L}_t\, \sigma_j^\alpha(t) \notag\\
&= -i\,[\sigma_j^\alpha(t), H_S(t)],
\end{align}
where we have neglected the contribution from $H_I$ compared to $H_S$.

The last term on the right-hand side of Eq.~(\ref{spin_in_bath}) is an operator in the environment space and corresponds to a fluctuating force. We define
\begin{equation}
\hat{\xi}_j^\alpha(t) = e^{it(1-\mathcal{P})\mathcal{L}}\, i(1-\mathcal{P})\mathcal{L} \sigma_j^\alpha, \qquad \hat{\xi}_{j}^{\alpha\dagger}(t) = \hat{\xi}_{j}^\alpha(t).
\end{equation}
Given that the environment possesses a large number of degrees of freedom and thus remains in thermal equilibrium, $\rho_B^{\mathrm{eq}} = Z_B^{-1} e^{-\beta H_B}$, the effect of the system on the environment can be neglected. Thus,
\begin{align}
\hat{\xi}_j^\alpha(t) &= e^{it(1-\mathcal{P})\mathcal{L}}\, i(1-\mathcal{P})\mathcal{L} \sigma_j^\alpha \notag\\
&\approx e^{it(1-\mathcal{P})\mathcal{L}_B} \left[ i(1-\mathcal{P})\mathcal{L} \sigma_j^\alpha \right] \notag\\
&= e^{it\mathcal{L}_B} \left[ i(1-\mathcal{P})\mathcal{L}_I \sigma_j^\alpha \right].
\end{align}
Assume $\operatorname{Tr}_B(\rho_B^{\mathrm{eq}} H_I) = 0$, then:
\begin{equation}
\langle \hat{\xi}_j^\alpha(t) \rangle_R = \operatorname{Tr}_B \left[ \rho_B^{\mathrm{eq}}\, e^{it\mathcal{L}_B}\, i(1-\mathcal{P})[H_I, \sigma_j^\alpha] \right] = 0.
\end{equation}

The second term on the right-hand side of Eq.~(\ref{spin_in_bath}), using the definition of $\hat{\xi}_j^\alpha(t)$, can be rewritten as
\begin{equation}
\int_0^t d\tau\, e^{i(t-\tau)\mathcal{L}}\, i\mathcal{P} \mathcal{L} \hat{\xi}_j^\alpha(\tau).
\end{equation}
Specifically,
\begin{align}
i\mathcal{P} \mathcal{L} \hat{\xi}_j^\alpha(\tau)
&= \sum_k \sum_{\mu\nu} \left(\sigma_k^\mu,\, i\mathcal{L}(1-\mathcal{P}) \hat{\xi}_j^\alpha(\tau) \right) (S_k^{-1})^{\nu\mu}\, \sigma_k^\nu \notag\\
&= -\sum_k \sum_{\mu\nu} \left(i\mathcal{L}(1-\mathcal{P})\sigma_k^\mu,\, \hat{\xi}_j^\alpha(\tau) \right) (S_k^{-1})^{\nu\mu}\, \sigma_k^\nu \notag\\
&= -\sum_k \sum_{\mu\nu} \left( \hat{\xi}_k^\mu(0),\, \hat{\xi}_j^\alpha(\tau) \right) (S_k^{-1})^{\nu\mu}\, \sigma_k^\nu.
\end{align}

Therefore, the entire second term becomes
\begin{align}
\int_0^t d\tau\, e^{i(t-\tau)\mathcal{L}}\, i\mathcal{P} \mathcal{L} \hat{\xi}_j^\alpha(\tau)
&= -\sum_{\mu,\nu} \int_0^t d\tau\, \left( \hat{\xi}_j^\mu(0),\, \hat{\xi}_j^\alpha(\tau) \right)\, (S_j^{-1})^{\nu\mu}\, \sigma_j^\mu(t-\tau) \notag\\
&= -\sum_\nu \int_0^t d\tau\, \Gamma_{\alpha\nu}^j(\tau)\, \sigma_j^\nu(t-\tau) \notag\\
&= -\sum_\nu \int_0^t d\tau\, \Gamma_{\alpha\nu}^j(t-\tau)\, \sigma_j^\nu(\tau),
\end{align}
where
\begin{equation}
\Gamma_{\alpha\nu}^j(\tau) \equiv \sum_\mu \left( \hat{\xi}_j^\mu(0),\, \hat{\xi}_j^\nu(\tau) \right)\, (S_j^{-1})^{\nu\mu}.
\end{equation}


Since the relaxation of the spin degrees of freedom is much faster than that of the spatial degrees of freedom, we employ the Markov approximation, so that $\Gamma_{\alpha\nu}^{j}(t-\tau) = \frac{1}{2}\gamma_{\alpha\nu}^{j} \delta(t-\tau)$. Therefore,
\begin{equation}
\int_{0}^{t} d\tau\, e^{i(t-\tau)\mathcal{L}}\, i\mathcal{P} \mathcal{L} \hat{\xi}_j^\alpha(\tau) = -\sum_{\nu} \gamma_{\alpha\nu} \sigma_{j}^{\nu}(t).
\end{equation}
Here, we assume that all baths are identical, so that $\gamma_{\alpha\nu}^{j} = \gamma_{\alpha\nu}$. The spectrum of the corresponding fluctuation force is therefore white noise,
\begin{align}
(\hat{\xi}_{j}^{\mu}(0),\, \hat{\xi}_{j}^{\alpha}(\tau)) &= \langle \hat{\xi}_{j}^{\mu\dagger}(0) \hat{\xi}_{j}^{\alpha}(\tau) \rangle_R \notag\\
&= \frac{1}{2} \sum_{\nu} \gamma_{\alpha\nu} \langle \sigma_{j}^{\mu} \sigma_{j}^{\nu} \rangle\, \delta(\tau).
\end{align}

Furthermore, if each component of the spin is independently dissipated, i.e., each of $\sigma_j^x,\, \sigma_j^y,\, \sigma_j^z$ is coupled to three independent and identical thermal baths (so that the dissipative spin dynamics of the $N$ particles involves $3N$ baths), we have
\begin{equation}
\rho_{B}^{\mathrm{eq}} = \prod_{\mu=1}^{3N} \rho_{\mu}^{\mathrm{eq}},
\end{equation}
where $\mu = (l, \alpha)$ denotes the bath coupled to the $\alpha$-th component of the $l$-th spin. Assume the interaction Hamiltonian is
\begin{equation}
H_I = \sum_l \sum_n \sum_{\alpha=1}^{3} a_{l\alpha}^{n}\, \hat{B}_{l\alpha}^{n} \sigma_l^\alpha,
\end{equation}
where $a$ are coefficients and $\hat{B}_{l\alpha}^n$ denotes the $n$-th degree of freedom of the bath coupled to the $\alpha$-th component of spin $l$. If $\operatorname{Tr}_B(\rho_B^{\mathrm{eq}}\, \hat{B}_{l\alpha}^n) = 0$, we have
\begin{equation}
\langle \hat{\xi}_j^{\mu \dagger}(0) \hat{\xi}_j^{\alpha}(\tau) \rangle = \delta_{\mu\alpha} \langle \hat{\xi}_j^{\alpha\dagger}(0) \hat{\xi}_j^{\alpha}(\tau) \rangle,
\end{equation}
and the result is independent of the index $j$. We then define
\begin{equation}
\gamma_s \equiv 2 \langle \hat{\xi}_j^{\alpha\dagger}(0) \hat{\xi}_j^{\alpha}(0) \rangle,
\end{equation}
and
\begin{equation}
\gamma_{\alpha\nu}^{j} = \gamma_s (S_j^{-1})^{\nu \alpha}.
\end{equation}
A straightforward calculation yields
\begin{align}
\gamma_{11}^j &= \gamma_{22}^j = \gamma_{33}^j = \gamma_s, \notag\\
\gamma_{12}^j &= -\gamma_{21}^j = i \sigma_j^{\mathrm{st}} \gamma_s, \notag\\
\gamma_{34}^j &=  -\gamma_s \sigma_j^{\mathrm{st}}.
\end{align}
Consequently, the equations of motion for the spin components take the form
\begin{align}
\frac{d}{dt} \sigma_j^x(t) &= -i [\sigma_j^x, H_S] - \gamma_s \sigma_j^x(t) - i \gamma_s \sigma_j^y(t)\, \sigma_j^{\mathrm{st}} + \hat{\xi}_j^x(t), \notag\\
\frac{d}{dt} \sigma_j^y(t) &= -i [\sigma_j^y, H_S] + i \gamma_s \sigma_j^{\mathrm{st}}\, \sigma_j^x(t) - \gamma_s \sigma_j^y(t) + \hat{\xi}_j^y(t), \notag\\
\frac{d}{dt} \sigma_j^z(t) &= -i [\sigma_j^z, H_S] - \gamma_s \left[ \sigma_j^z(t) - \sigma_j^{\mathrm{st}} \right] + \hat{\xi}_j^z(t).
\end{align}

Since $\sigma^x = i \sigma^y \sigma^z$, for small deviations from the steady state, one can replace $\sigma_z$ by its steady-state value $\sigma^{\mathrm{st}}$. Thus,
\begin{equation}
\sigma^x \approx i \sigma^y \sigma^{\mathrm{st}}, \qquad \sigma^y \approx -i \sigma^{\mathrm{st}} \sigma^x.
\end{equation}
With this approximation, the equations of motion become
\begin{align}
\frac{d}{dt} \sigma_j^x(t) &= -i [\sigma_j^x, H_S] - 2 \gamma_s \sigma_j^x(t) + \hat{\xi}_j^x(t), \notag\\
\frac{d}{dt} \sigma_j^y(t) &= -i [\sigma_j^y, H_S] - 2 \gamma_s \sigma_j^y(t) + \hat{\xi}_j^y(t), \notag\\
\frac{d}{dt} \sigma_j^z(t) &= -i [\sigma_j^z, H_S] - \gamma_s \left[ \sigma_j^z(t) - \sigma_j^{\mathrm{st}} \right] + \hat{\xi}_j^z(t).
\end{align}
It can be seen that the transverse relaxation rate (for $\sigma^x$ and $\sigma^y$) is faster than the longitudinal relaxation rate (for $\sigma^z$):
\begin{align}
T_2 = (2 \gamma_s)^{-1}, \qquad
T_1 = \gamma_s^{-1}.
\end{align}
where $T_2$ and $T_1$ are the transverse and longitudinal relaxation times, respectively. 
This analysis also shows that the transverse relaxation receives contributions from two sources: in addition to the usual environmental dissipation (corresponding to $T_1$), there is an additional contribution from phase randomization, i.e., the degeneration to the classical limit $\sigma_z \rightarrow \sigma^{\mathrm{st}}$.


In summary, the spin evolution equation can be written as
\begin{equation}
\frac{d}{dt} \vec{S}_j(t) = -i [\vec{S}_j, H_S] - \hat{\Gamma}_S \cdot (\vec{S}_j - \vec{S}_j^{(\mathrm{st})}) + \hat{\vec{F}}_{Sj}(t).
\end{equation}
If the steady-state value $\vec{S}_j^{(\mathrm{st})}$ is chosen along the $z$-axis, i.e., $\vec{S}_j^{(\mathrm{st})} = \frac{\hbar}{2} \sigma_j^{\mathrm{st}} \hat{e}_z$, then the dissipation tensor is $\hat{\Gamma}_S = \operatorname{diag} (2\gamma_S, 2\gamma_S, \gamma_S)$ and the noise term is
\begin{equation}
\hat{\vec{F}}_{Sj}(t) = \frac{\hbar}{2} \hat{\vec{\xi}}_j(t).
\end{equation}

The fluctuation-dissipation relation reads
\begin{align}
\gamma_s &\equiv 2 \langle \hat{\xi}_j^{\alpha\dagger}(t) \hat{\xi}_j^{\alpha}(0) \rangle, \notag\\
\langle \hat{F}_{Sj}^{\alpha\dagger}(t) \hat{F}_{Sj}^{\alpha'}(t') \rangle &= \frac{\hbar^2}{8} \gamma_s\, \delta_{ij} \delta_{\alpha\alpha'} \delta(t-t').
\end{align}

\subsection*{2. Spatial Dynamics}

We now turn to the dissipative dynamics of the spatial degrees of freedom. We assume that each spatial degree of freedom of every particle is coupled to its own independent, identical thermal bath, so that in total there are $3N$ independent quantum harmonic oscillator baths. Each bath is assumed to be sufficiently large so that its state is not affected by the system. The total Hamiltonian can be written as
\begin{equation}
H_T = H_S + H_E + V_I,
\end{equation}
where $H_S$ is the same as defined previously, and similarly, the spin degrees of freedom are treated as parameters in this part. The environment Hamiltonian is given by
\begin{equation}
H_E = \sum_{n=1}^N \sum_{\alpha=1}^3 h_{n\alpha}^E, \qquad
h_{n\alpha}^E = \sum_{\mu} \left[ \frac{1}{2} m_\mu^{n\alpha} (\omega_{\mu}^{n\alpha})^2 (\hat{q}_{\mu}^{n\alpha})^2 + \frac{(\hat{k}_{\mu}^{n\alpha})^2}{2m_\mu^{n\alpha}} \right].
\end{equation}
The equilibrium density matrix for the baths is
\begin{equation}
\rho_E = \prod_{n=1}^N \prod_{\alpha=1}^3 \rho_{\text{eq}}^{n\alpha}, \qquad
\rho_{\text{eq}}^{n\alpha} = \frac{1}{Z} \exp\left\{ - \beta \sum_{\mu} \left[ \frac{1}{2} m_\mu^{n\alpha} (\omega_{\mu}^{n\alpha})^2 (\hat{q}_{\mu}^{n\alpha})^2 + \frac{(\hat{k}_{\mu}^{n\alpha})^2}{2m_\mu^{n\alpha}} \right] \right\}.
\end{equation}
The interaction term is given by
\begin{align}
V_I &= \sum_{n=1}^N \sum_{\alpha=1}^3 x_{n\alpha} \sum_{\mu} C_{\mu}^{n\alpha} q_{\mu}^{n\alpha} 
+ \frac{1}{2} \sum_{n=1}^N \sum_{\alpha=1}^3 x_{n\alpha}^2 \sum_{\mu} \frac{(C_{\mu}^{n\alpha})^2}{m_\mu^{n\alpha} (\omega_\mu^{n\alpha})^2}.
\end{align}

We now consider the dissipation for each spatial degree of freedom separately. For the $\alpha$-th coordinate of the $n$-th particle, $x_{n\alpha}$, which is coupled to a single bath, we omit the $n\alpha$ subscript on bath quantities for simplicity. The equations of motion are
\begin{equation}
\begin{cases}
m \ddot{x}_{n\alpha} = -i [p_{n\alpha}, H_S] - \sum_{\mu} \frac{C_{\mu}^2}{m_{\mu} \omega_{\mu}^2} x_{n\alpha} + \sum_{\mu} C_{\mu} q_{\mu}, \\
m_{\mu} \ddot{q}_{\mu} = -m_{\mu} \omega_{\mu}^2 q_{\mu} + C_{\mu} x_{n\alpha},
\end{cases}
\end{equation}
where $p_{n\alpha}$ is the $\alpha$-th component of the momentum of the $n$-th particle.

Solving the equations of motion for the bath coordinates $q_{\mu}$ and substituting back into the equation for $x_{n\alpha}$, we obtain
\begin{align}
m \ddot{x}_{n\alpha}(t) = &-i [p_{n\alpha}, H_S] 
+ \sum_{\mu} C_\mu \left( q_\mu \cos \omega_\mu t + \frac{k_\mu}{m_\mu \omega_\mu} \sin \omega_\mu t \right) \notag \\
&- x_{n\alpha}(0) \sum_{\mu} \frac{C_\mu^2}{m_\mu \omega_\mu^2} \cos \omega_\mu t \notag\\
&- \sum_{\mu} \frac{C_\mu^2}{m_\mu \omega_\mu^2} \int_0^t dt' \cos \omega_\mu (t-t') \dot{x}_{n\alpha}(t').
\end{align}
Here $q_\mu$ and $k_\mu$ denote the initial values of the bath degrees of freedom.

Define the memory kernel and fluctuating force as
\begin{align}
\gamma(t-t') &\equiv \frac{1}{m} \sum_{\mu} \frac{C_\mu^2}{m_\mu \omega_\mu^2} \cos \omega_\mu (t-t') \Theta(t-t'), \\
\hat{f}^{\alpha}_{n}(t) &\equiv \frac{1}{m} \sum_{\mu} C_\mu \left( \hat{q}_\mu \cos \omega_\mu t + \frac{k_\mu}{m_\mu \omega_\mu} \sin \omega_\mu t \right),
\end{align}
where $\langle \hat{f}^{\alpha}_{n}(t) \rangle_R=Tr_E(\rho_E\hat{f}^{\alpha}_{n}(t)) = 0$.

The equation of motion then takes the form
\begin{equation}
\ddot{x}_{n\alpha}(t) = -\frac{i}{m} [p_{n\alpha}, H_S] -  \int_0^t d t'\, \gamma (t-t') \dot{x}_{n\alpha}(t') + \hat{f}_{n\alpha}(t), \qquad t > 0.
\end{equation}

The spectral density of the system-bath coupling is defined as
\begin{equation}
J(\omega) = \frac{\pi}{2} \sum_{\mu} \frac{C_\mu^2}{m_\mu \omega_\mu} \delta (\omega_\mu - \omega).
\end{equation}
By choosing an Ohmic spectral density, $J(\omega) = m \gamma \omega$, we have $\gamma(t-t') = 2 \gamma \delta(t-t')$. Therefore, the equation of motion simplifies to
\begin{equation}
\ddot{x}_{n\alpha}(t) = -\frac{i}{m} [p_{n\alpha}, H_S] - \gamma \dot{x}_{n\alpha}(t) + \hat{f}_{n\alpha}(t),
\end{equation}
with the correlation function of the noise following dissipation-fluctuation relation
\begin{align}
\frac{1}{2} \langle \{ \hat{f}_{n\alpha}(t), \hat{f}_{n\alpha}(t') \} \rangle_R 
&= \frac{1}{2} \langle \hat{f}_{n\alpha}(t) \hat{f}_{n\alpha}(t') + \hat{f}_{n\alpha}(t') \hat{f}_{n\alpha}(t) \rangle \notag\\
&= \frac{\hbar}{2m^2} \sum_{\mu} \frac{C_\mu^2}{m_\mu \omega_\mu} (2 n_\mu + 1) \cos \omega_\mu (t-t')\equiv \Phi(t-t'),
\end{align}
where $n_\mu$ is the Bose-Einstein distribution for the bath modes,
\begin{equation}
n_\mu = \frac{1}{e^{\beta \hbar \omega_\mu} - 1}.
\end{equation}

In the high temperature limit, the noise correlation simplifies to the classical dissipation-fluctuation relation
\begin{equation}
\langle \hat{f}_{n\alpha}(t) \hat{f}_{n\alpha}(t') \rangle_R \approx \frac{2\gamma k_B T}{m} \delta(t-t').
\end{equation}

\section{SII. Derivation of hydrodynamic equations}

In this section, we derive the hydrodynamic equations related to the microscopic QVM by kinetic approach. Our start point is an equivalent description of the stochastic active dynamic equations Eq.$\, (5)$ and Eq.$\, (6)$ in the overdamped regime (in the main text), i.e., the evolution equation of the distribution $f(\Vec{r},\Vec{v},t)$ in the single-particle phase space. Here, the overdamped condition eliminates the freedom $\Vec{S}$ approximately by the fact $\Vec{v}_j=u\Vec{S}_j$ on average. For simplicity, we assume that the randomness in the evolution of $f$ originates from the environment only. Therefore, the motion of $f$ satisfies\cite{de2013non,balescu1997statistical,PhysRevE.91.033303,tokatly2000hydrodynamics}
\begin{equation}
\elabel{Boltzmann_equation}
\partial_t f +\Vec{v}\cdot\nabla_r f + \Vec{a}\cdot\nabla_v f =I[f],
\end{equation}
where $\partial_t$, \(\nabla_r\) and \(\nabla_v\) respectively denote the differential operators with respect to time $t$, \(\Vec{r}\) and \(\Vec{v}\) as in conventional kinetic approach. \(I[f]\), following the Fokker-Planck theory \cite{van1992stochastic,risken1996fokker}, is given by $I[f]=\gamma\nabla_v\cdot(\Vec{v}-V(\Vec{r},t)+2\nabla_v/m\beta)f$. The term \(m\Vec{a}\) is the mean force acting on each particle which equals to the gradient of the mean single-particle free energy of the stationary state. 

We next determine this effective force exerted on per particle when the system is in the symmetry-broken phase. As mentioned in the main text, under the mean-field approximation, the energy of the system is 
\begin{equation}
\elabel{energy}
H_{ov}=E_0[N]+\sum_{j=1}^{N}[\frac{mv_j^2}{2}-\frac{J}{2u^2}N'\Vec{V}\cdot\Vec{v}_j],
\end{equation}
with the constraint $|\Vec{v}_j|=u$. For a small cell $\Delta V_r$ (while microscopically large enough), supposing both density $\rho(\Vec{r})$ and velocity $\Vec{V}(\Vec{r})$ are uniform in $\Delta V_r$ and denoting $N_r=\rho(\Vec{r})\Delta V_r$, the total free energy for the particles of $\Delta V_r$ is obtained by calculating
\begin{align}
\elabel{free_energy}
\mathrm{exp}[-\beta F(\Vec{r})\Delta V_r]&=\frac{(\Delta V_r)^{N_r}}{N_r!}\int\prod d\Vec{v}_j\mathrm{exp}[-\beta\sum_{i=1}^{N_r}(\frac{1}{2}mv_i^2-\frac{J}{2u^2}N'\Vec{V}(\Vec{r})\cdot\Vec{v}_i)]\delta(\Hat{v}_j-\Hat{V})\mathrm{exp}[-\kappa(v_j^2-u^2)]\notag \\
&=[(\frac{\Delta V_r}{N_r})\sqrt{\frac{2\pi}{\Tilde{m}\beta}}(\frac{J^2N'^2}{4\Tilde{m}^2u^4}V^2+\frac{1}{\Tilde{m}\beta})\mathrm{exp}(\beta\frac{J^2N'^2}{8\Tilde{m}^2u^4}V^2+\kappa u^2)]^{N_r},
\end{align}
and  in the high temperature limit, namely, when the average velocity of particles is much smaller than that of thermal motion, to the second order of $\beta mV^2$, the resulting free energy density is 
\begin{equation}
\elabel{free_energy}
F(\Vec{r}) =\rho( -\lambda|\Vec{V}|^2 +\eta|\Vec{V}|^4+F_0),
\end{equation}
where $\lambda=3J^2N'^2/8\Tilde{m}u^4$, $\eta=2\lambda^2\beta/9$ and $F_0=(3\ln{\Tilde{m}\beta}-\ln{2\pi}+2\ln\rho+2\kappa u^2)/2\beta$. Here, $\kappa$ is the Lagrange multiplier resulting from the constraint on the magnitude of velocities, and the renormalized mass is defined as $\Tilde{m} \equiv m + 2\kappa/\beta$. For such $\phi^4$ field, the symmetry spontaneously breaks at $|\Vec{V}_0|^2=9/(4\lambda\beta)$, which is just the order parameter\cite{landau2013statistical}. Thus, the mean force per particle is given by $m\Vec{a}=-\nabla_r(F/\rho)$. 

After obtaining the full evolution of $f(\Vec{r},\Vec{v},t)$, the hydrodynamic equations can be derived straightforwardly by definitions, namely,
\begin{equation}
\elabel{density}
\partial_t\rho=-\nabla_r(\rho\Vec{V}),
\end{equation}
\begin{equation}
\elabel{velocity_field}
\rho(\partial_t+\Vec{V}\cdot\nabla_r)\Vec{V}=-\nabla_r\cdot\Hat{P}+\rho \Vec{a},
\end{equation}
where $\Hat{P}=m\int d\Vec{v}\Vec{U}\Vec{U}f(\Vec{r},\Vec{v},t)$ is the stress tensor with $\Vec{U}\equiv\Vec{v}-\Vec{V}(\Vec{r},t)$ representing heat motion. As discussed in the main text, when $t$ is large, the direction of velocity and position are approximately the same, i.e., $\Vec{V}=\xi \Vec{r}$.   In this case, the dyadic $\nabla_r\Vec{V}$ on the right side of the equation is approximately proportional to the identity matrix and the force can be simplified as $m\Vec{a}=\Tilde{\lambda} \Vec{V}-\Tilde{\eta}\lvert \Vec{V}\rvert^2\Vec{V}$ with $\Tilde{\lambda}\equiv 2\xi \lambda$ and $\Tilde{\eta}=4\xi \eta$. The next order of acceleration simply corrects the nonlinear term. If we only consider the leading contribution, the result is that the coefficient of the nonlinear term changes from $1$ to $\lambda_1\equiv 1-2\lambda$.

Apparently, one still needs to solve $f(\Vec{r},\Vec{v},t)$ first to get the stress tensor. In the overdamped regime where $\gamma$ is very large, the evolution of $f$ \Eref{Boltzmann_equation} becomes approximately
\begin{equation}
\elabel{Boltzmann_equation2}
\partial_t f +\Vec{v}\cdot\nabla_r f + \Vec{a}\cdot\nabla_v f =-\gamma(f-f^{(0)})+R(\Vec{r},\Vec{v},t),
\end{equation}
where 
\begin{equation}
\elabel{f^(0)}
f^{(0)}(\Vec{v};\Vec{r},t)=\rho(\Vec{r},t)(\frac{m\beta}{2\pi})^{\frac{3}{2}}\mathrm{exp}[-\frac{m\beta}{2}(\Vec{v}-\Vec{V}(\Vec{r},t))^2]
\end{equation}
is stationary state corresponding to the Fokker-Planck operator $I[f]$, and $R(\Vec{r},\Vec{v},t)$ is the Gaussian random force with variance $\langle R(\Vec{r},\Vec{v},t)R(\Vec{r}',\Vec{v}',t')\rangle=2\Gamma\delta(\Vec{r}-\Vec{r}')\delta(\Vec{v}-\Vec{v}')\delta(t-t')$, which includes other impact of the environment. Here, the noise strength $\Gamma$ could be determined as follows. On the one hand, one can calculate the equal-time correlation of $\delta f=f-\langle f\rangle$ by \Eref{Boltzmann_equation2} and the result is $\langle \delta f(\Vec{r},\Vec{v},t)\delta f(\Vec{r}',\Vec{v}',t)\rangle=(\gamma)^{-1}\Gamma\delta(\Vec{r}-\Vec{r}')\delta(\Vec{v}-\Vec{v}')$. On the other hand, the equal-time correlation of $\delta f=f-\langle f\rangle$ could be approximated by the result in stationary state $f^{(0)}$, which is, $\langle \delta f(\Vec{r},\Vec{v},t)\delta f(\Vec{r}',\Vec{v}',t)\rangle\approx f^{(0)}\delta(\Vec{r}-\Vec{r}')\delta(\Vec{v}-\Vec{v}')$. Comparing these two results, one obtains $\Gamma=\gamma f^{(0)}$.

On average, up to the first order of $\gamma^{-1}$, the approximate solution of $f$ is 
\begin{align}
\elabel{f^(1)}
f^{(1)}(\Vec{v};\Vec{r},t)&=[1-\frac{1}{\gamma}(\partial_t+\Vec{v}\cdot\nabla_r+\Vec{a}\cdot\nabla_v)]f^{(0)}(\Vec{v};\Vec{r},t)\notag\\
&=[1-\frac{1}{\gamma}(-\nabla_r\cdot\Vec{V}(\Vec{r},t)+m\beta\Vec{U}\Vec{U}:\nabla_r\Vec{V}]f^{(0)}(\Vec{v};\Vec{r},t).
\end{align}
Here, the contraction of two tensors is defined as $\Vec{U}\Vec{U}:\nabla_r\Vec{V}=U_iU_j\partial_jV_i$ where the Einstein summation convention is applied. Hence, the resulting stress tensor is 
\begin{equation}
\elabel{stress_tensor}
\Hat{P}^{(1)}_{ab}=\frac{\rho}{\beta}\delta_{ab}-\frac{\rho}{m\gamma\beta}(\frac{\partial V_b}{\partial r_a}+\frac{\partial V_a}{\partial r_b})
\end{equation}
with $a,b$ being Cartesian indices.
In addition, the random term $R(\Vec{r},\Vec{v},t)$ in \Eref{Boltzmann_equation2} will give rise to a random force $\Vec{R}_v(\Vec{r},t)=\int \Vec{v}R(\Vec{r},\Vec{v},t) d\Vec{v} $ in the evolution equation of the velocity field $\Vec{V}(\Vec{r},t)$, whose covariance is easily obtained as $\langle R_{va}(\Vec{r},t) R_{vb}(\Vec{r}',t') \rangle_R = 2\gamma m\rho^{-1}[V_aV_b+\delta_{ab}(m\beta)^{-1}]\delta_{r_c} (\Vec{r}-\Vec{r}')\delta (t-t')$. Here $\delta_{r_c} (\Vec{r}-\Vec{r}')$ is the Dirac-$\delta$ like function implying the existence of the cutoff in spatial freedom. At last, we get the full hydrodynamic equation of $\Vec{V}(\Vec{r},t)$ \cite{de2013non, balescu1997statistical, PhysRevE.91.033303,tokatly2000hydrodynamics}, namely,
\begin{align}
\elabel{hydrodynamic_equation2}
m\left(\partial_t +\lambda_1 \Vec{V} \cdot \nabla_r \right) \Vec{V} = \Tilde{\lambda} \Vec{V}-\Tilde{\eta}\lvert \Vec{V}\rvert^2\Vec{V} - \frac{1}{ \rho} \nabla_r \cdot \Hat{P}+\Vec{R}_v,
\end{align}
as shown in Eq. $(10)$ in the main text.

\section{SIII. Analysis of the critical exponents}
In this part, we analyze the critical behavior of these equations by dynamic renormalization group method and calculate related critical exponents \cite{ma1975critical,ma2018modern,forster1977large}.

Firstly, we shift to a comoving coordinate frame moving with $\Vec{V}_0$, $\Vec{V}'=(\Vec{V}_\perp,V_\parallel \hat{e}_\parallel)$ with $\hat{e}_\parallel=\Vec{V}_0/\lvert \Vec{V}_0\rvert$. Here, $\Vec{V}_\perp(\Vec{r},t)$ is the Goldstone mode, which dominates the critical behavior of this system. $V_\parallel$ is fast mode, whose evolution follows approximately as $\partial_tV_\parallel=-\Tilde{\eta}(2V_\parallel+V_\perp^2)$. Obviously, the fast mode $V_\parallel\sim-V_\perp^2/2$ is a higher order infinitesimal, and thus can be ignored. As a result, the effective hydrodynamic equations then become,

\begin{align}
\elabel{hydrodynamic_equation3}
\left(\frac{\partial}{\partial t} + \lambda_1 \Vec{V}_\perp \cdot \Vec{\partial}_\perp \right) \Vec{V}_\perp 
= -B \Vec{\partial}_\perp \delta \rho + D \Vec{\partial}_\perp \Vec{\partial}_\perp \cdot \Vec{V}_\perp  + D \nabla_r^2 \Vec{V}_\perp + \Vec{R}_\perp,
\end{align}

\begin{align}
\elabel{hydrodynamic_equation4}
\partial_t\delta\rho=-A_I\Vec{\partial}_\perp\cdot\Vec{V}_\perp,
\end{align}
where $\Vec{\partial}_\perp$ denotes the directional derivative along $\Vec{V}_\perp$, $\delta\rho=\rho-\rho_0$ with $\rho_0$ being the mean of the local number density and $B\equiv(\rho_0\beta)^{-1}$. $D\equiv(\gamma m\beta)^{-1}$ is the diffusion coefficient, which also is subjected to the fluctuation-dissipation theorem (FDT). Notably, $\Vec{R}_\perp$ is the transformed random force with the covariance $\langle R_{\perp a}(\Vec{r},t) R_{\perp b}(\Vec{r}',t') \rangle_R = 2\Delta \Hat{I}_{ab}\delta_{r_c} (\Vec{r}-\Vec{r}')\delta_{\tau_c} (t-t')$ with the strength $\Delta\equiv \gamma(\rho_0 \beta)^{-1}$. At last, we emphasize that $A_I$ is the renormalized coefficient including the impact of the higher order (as well as nonlinear) term  $\Vec{\partial}_\perp(\delta\rho\Vec{V}_\perp)$.

Following the standard procedure of the dynamic renormalization group, we next eliminate the small wavelength components of $\delta \rho$ and $\Vec{V}_\perp$. This step is achieved easier in Fourier space. Namely, we take the Fourier transform 
\begin{align}
\rho(\Vec{k},\omega)=\int d\Vec{r} dt \delta\rho(\Vec{r},t)e^{i(\omega t-\Vec{k}\cdot\Vec{r})},
\end{align}
\begin{align}
\Vec{V}(\Vec{k},\omega)=\int d\Vec{r} dt \Vec{V}_\perp(\Vec{r},t)e^{i(\omega t-\Vec{k}\cdot\Vec{r})},
\end{align}
\begin{align}
\Vec{R}(\Vec{k},\omega)=\int d\Vec{r} dt \Vec{R}_\perp(\Vec{r},t)e^{i(\omega t-\Vec{k}\cdot\Vec{r})}.
\end{align}
The variance of the random force becomes $\Vec{R}(\Vec{k},\omega)$ is $\langle \Vec{R}(\Vec{k},\omega) \Vec{R}(\Vec{k}',\omega') \rangle_R = (2\pi)^4\Delta\Hat{I}\delta(\omega-\omega')\delta(\Vec{k}-\Vec{k}')$ with $\Delta=\Delta(1-e^{-k^2/\Lambda^2})$. Here, $\Lambda$ is the ultraviolet cutoff of $k$, which is the mathematical meaning of the above Dirac-$\delta$ like function. The transformed hydrodynamic equations are
\begin{align}
\elabel{hydrodynamic_equation6}
[-i\omega +D(k^2+\Vec{k}_\perp\Vec{k}_\perp)]\cdot\Vec{V}(\Vec{k},\omega)=\Vec{R}(\Vec{k},\omega)-iB\Vec{k}_\perp\rho(\Vec{k},\omega)-i\lambda_1\int_{q,\Omega}\Vec{V}(\Vec{k}-\Vec{q},\omega-\Omega)\cdot\Vec{q}_\perp\Vec{V}(\Vec{q},\Omega)
\end{align}
and
\begin{align}
\elabel{hydrodynamic_equation7}
i\omega\rho(\Vec{k},\omega)=iA_I\Vec{k}_\perp\cdot\Vec{V}(\Vec{k},\omega).
\end{align}
Combining these two equations, we obtained \Eref{hydrodynamic_equation5} of the main text, i.e.,
\begin{align}
\elabel{hydrodynamic_equation5}
[-i\omega+i\frac{A_IB}{\omega}\Vec{k}_\perp\Vec{k}_\perp +D(k^2+\Vec{k}_\perp\Vec{k}_\perp)]\cdot\Vec{V}(\Vec{k},\omega)=\Vec{R}(\Vec{k},\omega)-i\lambda_1\int_{q,\Omega}\Vec{V}(\Vec{k}-\Vec{q},\omega-\Omega)\cdot\Vec{q}_\perp\Vec{V}(\Vec{q},\Omega).
\end{align}
Defining the propagator as
\begin{align}
\elabel{G}
\Hat{G}(\Vec{k},\omega)=[-i\omega+D(k^2+\Vec{k}_\perp\Vec{k}_\perp)+i\frac{A_IB}{\omega}\Vec{k}_\perp\Vec{k}_\perp]^{-1}.
\end{align}
The formal solution of $\Vec{V}(\Vec{k},\omega)$ could be written as
\begin{align}
\elabel{solution_V}
\Vec{V}(\Vec{k},\omega)=\Hat{G}(\Vec{k},\omega)\cdot\Vec{R}(\Vec{k},\omega)-i\lambda_1 \Hat{G}(\Vec{k},\omega)\cdot\int^\Lambda \frac{d\Vec{k}'}{(2\pi)^3}\int^{+\infty}_{-\infty}\frac{d\omega'}{2\pi}\Vec{V}(\Vec{k}-\Vec{k}',\omega-\omega')\cdot\Vec{k}'_\perp\Vec{V}(\Vec{k}',\omega'),
\end{align}
where $\int^\Lambda \frac{d\Vec{k}'}{(2\pi)^3}$ indicates the upper limit of the wave number is $\Lambda$. For simplicity, we denote $\int^\Lambda \frac{d\Vec{k}'}{(2\pi)^3}\int^{+\infty}_{-\infty}\frac{d\omega'}{2\pi}\equiv \int_{k',\omega'}$, and $\int^<_{k',\omega'}$ means the integral interval of wave number is $0<k'<\Lambda/s$, while$\int^>_{k',\omega'}$ means the integral interval of wave number is the shell $\Lambda/s<k'<\Lambda$. The first step thus becomes to project the equations of motion onto the phase space spanned by modes $\Vec{V}_s(\Vec{k},\omega)$ with $0<k<\Lambda/s$, and pushes the remainder into the appropriately redefined noise, namely, eliminating those modes $\Vec{V}_f(\Vec{q},\Omega)$ with short wavelength $\Lambda/s<q<\Lambda$. This is done by formally solving the equation for $\Vec{V}_f(\Vec{q},\Omega)$, and then substituting the solution into the equation for $\Vec{V}_s(\Vec{k},\omega)$ to eliminate their explicit dependence on $\Vec{V}_f(\Vec{q},\Omega)$, and finally averaging over the part of the random force $\Vec{R}(\Vec{q},\Omega)$ that acts in the shell $\Lambda/s<q<\Lambda$.

To further discuss, it is convenient and necessary to introduce the zero-order velocity spectrum, that is, the result when ignoring the nonlinear $\lambda_1$-term. Denote $\Vec{V}^{(0)}(\Vec{k},\omega)$ as the solution for velocity when ignoring the nonlinear term. The correlation between $\Vec{V}^{(0)}(\Vec{k},\omega)$ and $\Vec{V}^{(0)}(\Vec{k}',\omega')$ is
\begin{align}
\elabel{correlation_V}
\langle \Vec{V}^{(0)}(\Vec{k},\omega) \Vec{V}^{(0)}(\Vec{k}',\omega') \rangle_R = (2\pi)^4\Hat{C}^{(0)}(\Vec{k},\omega)\delta(\omega-\omega')\delta(\Vec{k}-\Vec{k}'),
\end{align}
where the zero-order velocity spectrum is given by
\begin{align}
\elabel{C}
\Hat{C}^{(0)}(\Vec{k},\omega)=\frac{2\omega^2\Delta}{D^2\omega^2(k^2+\Vec{k}_\perp\Vec{k}_\perp)^2+(\omega^2-A_IB\Vec{k}_\perp\Vec{k}_\perp)^2}.
\end{align}
Besides, the following quantities and relations are also quite useful,

\begin{itemize}
    \item[a.]  the real part of the propagator $\Hat{G}(\Vec{k},\omega)$ is
\begin{align}
\elabel{G_1}
\Hat{G_1}(\Vec{k},\omega)
=\frac{D(k^2+\Vec{k}_\perp\Vec{k}_\perp)\omega^2}{D^2\omega^2(k^2+\Vec{k}_\perp\Vec{k}_\perp)^2+(\omega^2-A_IB\Vec{k}_\perp\Vec{k}_\perp)^2},
\end{align}
which is the even function with respect to both $\omega$ and $\Vec{k}$;

    \item[b.] the imaginary part of the propagator $\Hat{G}(\Vec{k},\omega)$ is
\begin{align}
\elabel{G_2}
\Hat{G_2}(\Vec{k},\omega)=\frac{\omega^3-A_IB\omega\Vec{k}_\perp\Vec{k}_\perp}{D^2\omega^2(k^2+\Vec{k}_\perp\Vec{k}_\perp)^2+(\omega^2-A_IB\Vec{k}_\perp\Vec{k}_\perp)^2},
\end{align}
which is the odd function with respect to $\omega$ while the even function with respect to $\Vec{k}$;

\item[c.] the zero-order velocity spectrum $\Hat{C}^{(0)}(\Vec{k},\omega)$ is the even function with respect to both $\omega$ and $\Vec{k}$.
\end{itemize}

With these preparations in place, we execute the first step. In general, the nonlinear term couples the desired short-wavelength modes $\Vec{V}_s(\Vec{k},\omega)$ with the unexpected long-wavelength modes $\Vec{V}_f(\Vec{q},\Omega)$. The conventional approach to tackle this is to write the iterative solution as a power series in $\lambda_1$. We retain up to $\lambda_1^2$ in the final result, which means the required solution for $\Vec{V}_f(\Vec{q},\Omega)$ is 
\begin{align}
\elabel{solution_V_f}
\Vec{V}_f(\Vec{q},\Omega)=\Hat{G}(\Vec{q},\Omega)\cdot\Vec{R}(\Vec{q},\Omega)-i\lambda_1 \Hat{G}(\Vec{q},\Omega)\cdot\int_{k',\omega'}\Vec{V}^{(0)}(\Vec{q}-\Vec{k}',\Omega-\omega')\cdot\Vec{k}'_\perp\Vec{V}^{(0)}(\Vec{k}',\omega').
\end{align}
Substituting this approximate solution into the iterative solution of slow mode $\Vec{V}_s(\Vec{k},\omega)$ and only keeping terms containing even numbers of $\Vec{V}_f^{(0)}$ (since the average of terms containing odd numbers of $\Vec{V}_f^{(0)}$ is zero), the result is 
\begin{align}
\elabel{solution_V_s}
\Vec{V}_s(\Vec{k},\omega)=&\Hat{G}(\Vec{k},\omega)\cdot\Vec{R}(\Vec{k},\omega)-i\lambda_1 \Hat{G}(\Vec{k},\omega)\cdot\int^>_{k',\omega'}\Vec{V}_f^{(0)}(\Vec{k}-\Vec{k}',\omega-\omega')\cdot\Vec{k}'_\perp\Vec{V}_f^{(0)}(\Vec{k}',\omega')\notag\\
&-i \lambda_1  \Hat{G}(\Vec{k},\omega)\cdot\int^<_{k',\omega'}\Vec{V}_s^{(0)}(\Vec{k}-\Vec{k}',\omega-\omega')\cdot\Vec{k}'_\perp\Vec{V}_s^{(0)}(\Vec{k}',\omega')\notag\\
&-\lambda_1^2\Hat{G}(\Vec{k},\omega)\cdot\int^<_{k',\omega'}(\Hat{G}(\Vec{k}-\Vec{k}',\omega-\omega')\cdot\int^<_{k'',\omega''}\Vec{V}_s^{(0)}(\Vec{k}-\Vec{k}'-\Vec{k}'',\omega-\omega'-\omega'')\cdot\Vec{k}''_\perp\Vec{V}_s^{(0)}(\Vec{k}'',\omega''))\cdot\Vec{k}'_\perp\Vec{V}_s^{(0)}(\Vec{k}',\omega')\notag\\
&-\lambda_1^2\Hat{G}(\Vec{k},\omega)\cdot\int^<_{k',\omega'}\Vec{V}_s^{(0)}(\Vec{k}-\Vec{k}',\omega-\omega')\cdot\Vec{k}'_\perp\Hat{G}(\Vec{k}',\omega')\cdot\int^<_{k'',\omega''}\Vec{V}_s^{(0)}(\Vec{k}'-\Vec{k}'',\omega'-\omega'')\cdot\Vec{k}''_\perp\Vec{V}_s^{(0)}(\Vec{k}'',\omega'')\notag\\
&-\lambda_1^2\Hat{G}(\Vec{k},\omega)\cdot\int^<_{k',\omega'}(\Hat{G}(\Vec{k}-\Vec{k}',\omega-\omega')\cdot\int^>_{k'',\omega''}\Vec{V}_f^{(0)}(\Vec{k}-\Vec{k}'-\Vec{k}'',\omega-\omega'-\omega'')\cdot\Vec{k}''_\perp\Vec{V}_f^{(0)}(\Vec{k}'',\omega''))\cdot\Vec{k}'_\perp\Vec{V}_s^{(0)}(\Vec{k}',\omega')\notag\\
&-\lambda_1^2\Hat{G}(\Vec{k},\omega)\cdot\int^>_{k',\omega'}(\Hat{G}(\Vec{k}-\Vec{k}',\omega-\omega')\cdot\int^<_{k'',\omega''}\Vec{V}_f^{(0)}(\Vec{k}-\Vec{k}'-\Vec{k}'',\omega-\omega'-\omega'')\cdot\Vec{k}''_\perp\Vec{V}_s^{(0)}(\Vec{k}'',\omega''))\cdot\Vec{k}'_\perp\Vec{V}_f^{(0)}(\Vec{k}',\omega')\notag\\
&-\lambda_1^2\Hat{G}(\Vec{k},\omega)\cdot\int^>_{k',\omega'}(\Hat{G}(\Vec{k}-\Vec{k}',\omega-\omega')\cdot\int^>_{k'',\omega''}\Vec{V}_s^{(0)}(\Vec{k}-\Vec{k}'-\Vec{k}'',\omega-\omega'-\omega'')\cdot\Vec{k}''_\perp\Vec{V}_f^{(0)}(\Vec{k}'',\omega''))\cdot\Vec{k}'_\perp\Vec{V}_f^{(0)}(\Vec{k}',\omega')\notag\\
&-\lambda_1^2\Hat{G}(\Vec{k},\omega)\cdot\int_{k',\omega'}\Vec{V}_s^{(0)}(\Vec{k}-\Vec{k}',\omega-\omega')\cdot\Vec{k}'_\perp(\Hat{G}(\Vec{k}',\omega')\cdot\int^>_{k'',\omega''}\Vec{V}_f^{(0)}(\Vec{k}'-\Vec{k}'',\omega'-\omega'')\cdot\Vec{k}''_\perp\Vec{V}_f^{(0)}(\Vec{k}'',\omega'')\notag\\
&-\lambda_1^2\Hat{G}(\Vec{k},\omega)\cdot\int_{k',\omega'}\Vec{V}_f^{(0)}(\Vec{k}-\Vec{k}',\omega-\omega')\cdot\Vec{k}'_\perp(\Hat{G}(\Vec{k}',\omega')\cdot\int^>_{k'',\omega''}\Vec{V}_s^{(0)}(\Vec{k}'-\Vec{k}'',\omega'-\omega'')\cdot\Vec{k}''_\perp\Vec{V}_f^{(0)}(\Vec{k}'',\omega'')\notag\\
&-\lambda_1^2\Hat{G}(\Vec{k},\omega)\cdot\int_{k',\omega'}\Vec{V}_f^{(0)}(\Vec{k}-\Vec{k}',\omega-\omega')\cdot\Vec{k}'_\perp(\Hat{G}(\Vec{k}',\omega')\cdot\int^<_{k'',\omega''}\Vec{V}_f^{(0)}(\Vec{k}'-\Vec{k}'',\omega'-\omega'')\cdot\Vec{k}''_\perp\Vec{V}_s^{(0)}(\Vec{k}'',\omega'').
\end{align}
Next, we need to average this expression over $\Vec{V}_f^{(0)}$. This procedure is a little bit sophisticated but quite straightforward. Using \Eref{correlation_V}, we obtain
\begin{align}
\elabel{solution_V_s_afterAverage}
\Vec{V}_s(\Vec{k},\omega)=
&\Hat{G}(\Vec{k},\omega)\cdot\Vec{R}(\Vec{k},\omega)
-i\lambda_1 \Hat{G}(\Vec{k},\omega)\cdot\int^>_{k',\omega'}\Vec{V}_f^{(0)}(\Vec{k}-\Vec{k}',\omega-\omega')\cdot\Vec{k}'_\perp\Vec{V}_f^{(0)}(\Vec{k}',\omega')\notag\\
&-i \lambda_1  \Hat{G}(\Vec{k},\omega)\cdot\int^<_{k',\omega'}\Vec{V}_s^{(0)}(\Vec{k}-\Vec{k}',\omega-\omega')\cdot\Vec{k}'_\perp\Vec{V}_s^{(0)}(\Vec{k}',\omega')\notag\\
&-\lambda_1^2\Hat{G}(\Vec{k},\omega)\cdot\int^<_{k',\omega'}(\Hat{G}(\Vec{k}-\Vec{k}',\omega-\omega')\cdot\int^<_{k'',\omega''}\Vec{V}_s^{(0)}(\Vec{k}-\Vec{k}'-\Vec{k}'',\omega-\omega'-\omega'')\cdot\Vec{k}''_\perp\Vec{V}_s^{(0)}(\Vec{k}'',\omega''))\cdot\Vec{k}'_\perp\Vec{V}_s^{(0)}(\Vec{k}',\omega')\notag\\
&-\lambda_1^2\Hat{G}(\Vec{k},\omega)\cdot\int^<_{k',\omega'}\Vec{V}_s^{(0)}(\Vec{k}-\Vec{k}',\omega-\omega')\cdot\Vec{k}'_\perp(\Hat{G}(\Vec{k}',\omega')\cdot\int^<_{k'',\omega''}\Vec{V}_s^{(0)}(\Vec{k}'-\Vec{k}'',\omega'-\omega'')\cdot\Vec{k}''_\perp\Vec{V}_s^{(0)}(\Vec{k}'',\omega'')\notag\\
&-(2\pi)^4\lambda_1^2\Hat{G}(\Vec{k},\omega)\cdot\int^>_{k',\omega'}\Vec{k}_\perp\cdot\Hat{G}(0,0)\cdot\Hat{C}_f^{(0)}(\Vec{k}',\omega')\cdot\Vec{k}'_\perp\Vec{V}_s^{(0)}(\Vec{k},\omega)\notag\\
&-(2\pi)^4\lambda_1^2\Hat{G}(\Vec{k},\omega)\cdot\int^>_{k',\omega'}\Hat{C}_f^{(0)}(\Vec{k}',\omega')\cdot\Vec{k}_\perp\Vec{k}'_\perp\cdot\Hat{G}(\Vec{k}-\Vec{k}',\omega-\omega')\cdot\Vec{V}_s^{(0)}(\Vec{k},\omega)\notag\\
&+(2\pi)^4\lambda_1^2\Hat{G}(\Vec{k},\omega)\cdot\int^>_{k',\omega'}\Hat{C}_f^{(0)}(\Vec{k}',\omega')\cdot\Hat{G}(\Vec{k}-\Vec{k}',\omega-\omega')\cdot\Vec{k}'_\perp\Vec{k}'_\perp\cdot\Vec{V}_s^{(0)}(\Vec{k},\omega)\notag\\
&-(2\pi)^4\lambda_1^2\Hat{G}(\Vec{k},\omega)\cdot\int^>_{k',\omega'}\Hat{G}(\Vec{k}+\Vec{k}',\omega+\omega')\cdot\Hat{C}_f^{(0)}(\Vec{k}',\omega')\cdot(\Vec{k}_\perp+\Vec{k}'_\perp)\Vec{k}'_\perp\cdot\Vec{V}_s^{(0)}(\Vec{k},\omega)\notag\\
&-(2\pi)^4\lambda_1^2\Hat{G}(\Vec{k},\omega)\cdot\int^>_{k',\omega'}\Vec{k}_\perp\cdot\Hat{C}_f^{(0)}(\Vec{k}-\Vec{k}',\omega-\omega')\cdot\Vec{k}'_\perp\Hat{G}(\Vec{k}',\omega')\cdot\Vec{V}_s^{(0)}(\Vec{k},\omega).
\end{align}
Now let us try to revert the solution \Eref{solution_V_s_afterAverage} to the form similar to \Eref{hydrodynamic_equation5} to find the renormalized parameters $D_I$, $\Delta_I$, $A_IB_I$ and $\lambda_I$. This is easily done by multiplying both sides of the equation by $\Hat{G}^{-1}$ and then moving the last four terms on the right side of the equation to the left side, the result reads
\begin{align}
\elabel{solution_V_s_reduction}
[-i\omega+i\frac{A_IB}{\omega}\Vec{k}_\perp\Vec{k}_\perp+D(k^2+\Vec{k}_\perp\Vec{k}_\perp)+\lambda_1^2I]\cdot\Vec{V}(\Vec{k},\omega)=&\Vec{R}(\Vec{k},\omega)-i\lambda_1 \int^>_{k',\omega'}\Vec{V}_f^{(0)}(\Vec{k}-\Vec{k}',\omega-\omega')\cdot\Vec{k}'_\perp\Vec{V}_f^{(0)}(\Vec{k}',\omega')\notag\\
&-i\lambda_1\int_{q,\Omega}\Vec{V}(\Vec{k}-\Vec{q},\omega-\Omega)\cdot\Vec{q}_\perp\Vec{V}(\Vec{q},\Omega).
\end{align}
Here, $\lambda_1^2I$ refer to the contribution from the last four terms on the right side of \Eref{solution_V_s_afterAverage},which is
\begin{align}
\elabel{I}
I=&+(2\pi)^4\int^>_{k',\omega'}\Hat{C}_f^{(0)}(\Vec{k}',\omega')\cdot\Vec{k}_\perp\Vec{k}'_\perp\cdot\Hat{G}(\Vec{k}-\Vec{k}',\omega-\omega')\notag\\
&-(2\pi)^4\int^>_{k',\omega'}\Hat{G}(\Vec{k}-\Vec{k}',\omega-\omega')\cdot\Hat{C}_f^{(0)}(\Vec{k}',\omega')\cdot\Vec{k}_\perp\Vec{k}'_\perp\notag\\
&-(2\pi)^4\int^>_{k',\omega'}\Hat{C}_f^{(0)}(\Vec{k}',\omega')\cdot\Hat{G}(\Vec{k}+\Vec{k}',\omega+\omega')\cdot\Vec{k}'_\perp\Vec{k}'_\perp\notag\\
&+(2\pi)^4\int^>_{k',\omega'}\Hat{G}(\Vec{k}+\Vec{k}',\omega+\omega')\cdot\Hat{C}_f^{(0)}(\Vec{k}',\omega')\cdot\Vec{k}'_\perp\Vec{k}'_\perp\notag\\
&+(2\pi)^4\int^>_{k',\omega'}\Vec{k}_\perp\cdot\Hat{C}_f^{(0)}(\Vec{k}-\Vec{k}',\omega-\omega')\cdot\Vec{k}'_\perp\Hat{G}(\Vec{k}',\omega'),
\end{align}
while the third and the forth terms on the right side of \Eref{solution_V_s_afterAverage} are absorbed into the last term of \Eref{solution_V_s_reduction}. 
Apparently, as shown in \Eref{solution_V_s_reduction}, the corrections to $D$ and $A_IB$ come from the real part and the imaginary part of $\lambda_1^2I$, respectively, and $-i\lambda_1 \int^>_{k',\omega'}\Vec{V}_f^{(0)}(\Vec{k}-\Vec{k}',\omega-\omega')\cdot\Vec{k}'_\perp\Vec{V}_f^{(0)}(\Vec{k}',\omega')$ is the correction to the random force. However, the nonlinear coefficient remains unchanged, i.e., $\lambda_I=\lambda_1$. Therefore, the next task is to calculate these corrections.

Noting that both $k$ and $\omega$ are very small, we can Taylor expand $\Hat{C}_f^{(0)}$s and $\Hat{G}$ in integrands of $I$ to the first order, and obviously, the integral result $I$ will be proportional to $k^2$. The outcome of the expanding reads
\begin{align}
I=&+(2\pi)^4\int^>_{k',\omega'}\Hat{C}_f^{(0)}(\Vec{k}',\omega')\cdot\Vec{k}_\perp\Vec{k}'_\perp\cdot\Hat{G}(-\Vec{k}',-\omega')\notag\\
&-(2\pi)^4\int^>_{k',\omega'}\Hat{G}(-\Vec{k}',-\omega')\cdot\Hat{C}_f^{(0)}(\Vec{k}',\omega')\cdot\Vec{k}_\perp\Vec{k}'_\perp\notag\\
&-(2\pi)^4\int^>_{k',\omega'}\Hat{C}_f^{(0)}(\Vec{k}',\omega')\cdot\Hat{G}(\Vec{k}',\omega')\cdot\Vec{k}'_\perp\Vec{k}'_\perp\notag\\
&+(2\pi)^4\int^>_{k',\omega'}\Hat{G}(\Vec{k}',\omega')\cdot\Hat{C}_f^{(0)}(\Vec{k}',\omega')\cdot\Vec{k}'_\perp\Vec{k}'_\perp\notag\\
&+(2\pi)^4\int^>_{k',\omega'}\Vec{k}_\perp\cdot\Hat{C}_f^{(0)}(\Vec{k}',\omega')\cdot\Vec{k}'_\perp\Hat{G}(\Vec{k}',\omega')\notag\\
&-(2\pi)^4\int^>_{k',\omega'}\Hat{C}_f^{(0)}(\Vec{k}',\omega')\cdot\Vec{k}_\perp\Vec{k}'_\perp\cdot(\Vec{k}\cdot\nabla_{k'}+\omega\partial_{\omega'})\Hat{G}(\Vec{k}',\omega')\notag\\
&+(2\pi)^4\int^>_{k',\omega'}(\Vec{k}\cdot\nabla_{k'}+\omega\partial_{\omega'})\Hat{G}(\Vec{k}',\omega')\cdot\Hat{C}_f^{(0)}(\Vec{k}',\omega')\cdot\Vec{k}_\perp\Vec{k}'_\perp\notag\\
&-(2\pi)^4\int^>_{k',\omega'}\Hat{C}_f^{(0)}(\Vec{k}',\omega')\cdot(\Vec{k}\cdot\nabla_{k'}+\omega\partial_{\omega'})\Hat{G}(\Vec{k}',\omega')\cdot\Vec{k}'_\perp\Vec{k}'_\perp\notag\\
&+(2\pi)^4\int^>_{k',\omega'}(\Vec{k}\cdot\nabla_{k'}+\omega\partial_{\omega'})\Hat{G}(\Vec{k}',\omega')\cdot\Hat{C}_f^{(0)}(\Vec{k}',\omega')\cdot\Vec{k}'_\perp\Vec{k}'_\perp\notag\\
&-(2\pi)^4\int^>_{k',\omega'}\Vec{k}_\perp\cdot(\Vec{k}\cdot\nabla_{k'}+\omega\partial_{\omega'})\Hat{C}_f^{(0)}(\Vec{k}',\omega')\cdot\Vec{k}'_\perp\Hat{G}(\Vec{k}',\omega').
\end{align}
By analyzing the parity of the integrands, we obtain
\begin{align}
I=&-(2\pi)^4\int^>_{k',\omega'}\Hat{C}_f^{(0)}(\Vec{k}',\omega')\cdot\Vec{k}_\perp\Vec{k}'_\perp\cdot\Vec{k}_\perp\cdot\Vec{\partial}'_{k\perp}\Hat{G}(\Vec{k}',\omega')\notag\\
&+(2\pi)^4\int^>_{k',\omega'}\Vec{k}_\perp\cdot\Vec{\partial}'_{k\perp}\Hat{G}(\Vec{k}',\omega')\cdot\Hat{C}_f^{(0)}(\Vec{k}',\omega')\cdot\Vec{k}_\perp\Vec{k}'_\perp\notag\\
&-(2\pi)^4\int^>_{k',\omega'}\Vec{k}_\perp\cdot\Vec{k}_\perp\cdot\Vec{\partial}'_{k\perp}\Hat{C}_f^{(0)}(\Vec{k}',\omega')\cdot\Vec{k}'_\perp\Hat{G}(\Vec{k}',\omega'),
\end{align}
where $\Vec{\partial}'_{k\perp}\equiv\partial/\partial\Vec{k}'_\perp$. Consequently, the correction to the diffusion coefficient $D$ comes from
\begin{align}
Re[I]=&-(2\pi)^4\int^>_{k',\omega'}\Hat{C}_f^{(0)}(\Vec{k}',\omega')\cdot\Vec{k}_\perp\Vec{k}'_\perp\cdot\Vec{k}_\perp\cdot\Vec{\partial}'_{k\perp}\Hat{G}_1(\Vec{k}',\omega')\notag\\
&+(2\pi)^4\int^>_{k',\omega'}\Vec{k}_\perp\cdot\Vec{\partial}'_{k\perp}\Hat{G}_1(\Vec{k}',\omega')\cdot\Hat{C}_f^{(0)}(\Vec{k}',\omega')\cdot\Vec{k}_\perp\Vec{k}'_\perp\notag\\
&-(2\pi)^4\int^>_{k',\omega'}\Vec{k}_\perp\cdot\Vec{k}_\perp\cdot\Vec{\partial}'_{k\perp}\Hat{C}_f^{(0)}(\Vec{k}',\omega')\cdot\Vec{k}'_\perp\Hat{G}_1(\Vec{k}',\omega'),
\end{align}
while $A_IB$, as a whole, is corrected by 
\begin{align}
\elabel{Iim}
Im[I]=&-(2\pi)^4\int^>_{k',\omega'}\Hat{C}_f^{(0)}(\Vec{k}',\omega')\cdot\Vec{k}_\perp\Vec{k}'_\perp\cdot\Vec{k}_\perp\cdot\Vec{\partial}'_{k\perp}\Hat{G}_2(\Vec{k}',\omega')\notag\\
&+(2\pi)^4\int^>_{k',\omega'}\Vec{k}_\perp\cdot\Vec{\partial}'_{k\perp}\Hat{G}_2(\Vec{k}',\omega')\cdot\Hat{C}_f^{(0)}(\Vec{k}',\omega')\cdot\Vec{k}_\perp\Vec{k}'_\perp\notag\\
&-(2\pi)^4\int^>_{k',\omega'}\Vec{k}_\perp\cdot\Vec{k}_\perp\cdot\Vec{\partial}'_{k\perp}\Hat{C}_f^{(0)}(\Vec{k}',\omega')\cdot\Vec{k}'_\perp\Hat{G}_2(\Vec{k}',\omega').
\end{align}
Interestingly, note that $\Hat{G}_2(\Vec{k}',\omega')$ is odd with respect to $\omega'$, so the integral result of \Eref{Iim} is zero, which means $A_IB$ does not have correction either. It should be emphasized that it will not arise conflict here since $A_I$ is already a renormalized value. In fact, one can conclude that $B$, measuring the response of the velocity field to changes in density, remains unchanged in new reduced hydrodynamic equation,i.e., $B_I=B$. 

We now estimate $\delta D_I$ as
\begin{align}
\delta D_I&\approx-(2\pi)^4\lambda_1^2\int^>_{k',\omega'}\Vec{\partial}'_{k\perp}\Hat{C}(\Vec{k}',\omega')\cdot\Vec{k}'_\perp\Hat{G}_1(\Vec{k}',\omega')\notag\\
&\approx-(2\pi)^4\lambda_1^2\int^>_{k',\omega'}\Vec{k}'\cdot\nabla_{k'}C(k',\omega')G_1(k',\omega')
\end{align}
with 
\begin{align}
\elabel{C}
C(k,\omega)=\frac{2\omega^2\Delta}{D^2\omega^2\mu^2 k^4+(\omega^2-A_IB\nu k^2)^2}
\end{align}
and
\begin{align}
\elabel{G}
G_1(k,\omega)=\frac{D\mu k^2\omega^2}{D^2\omega^2\mu^2 k^4+(\omega^2-A_IB\nu k^2)^2}
\end{align}
where we approximate $k^2+k_jk_j~(1+\nu)k^2=\mu k^2$. $\mu$, as well as $\nu$, could be regarded as a structure factor, which does not influence how the final integral result depends on $s$. Therefore, $\delta D_I$ is estimated as 
\begin{align}
\delta D_I&\approx D\Bar{\lambda}_1^2 F(s;\Lambda,\frac{A_IB}{D})\notag\\
&\approx D\Bar{\lambda}_1^2(F_{31}(\Lambda)(s-1)+\frac{AB}{D}F_{33}(\Lambda)(s^3-1)+\cdots),
\end{align}
with $\Bar{\lambda}_1^2\equiv\Delta\lambda_1^2/D^3$. Here $F(s;\Lambda,A_IB/D)$ is a polynomial function of $s$, which contains only odd powers terms and the lowest order is $s^1$. The expanded coefficients $F_{31}$ and $F_{33}$ are related to $\Lambda$.

The last piece is the correction to the noise strength devoted by $-i\lambda_1 \int^>_{k',\omega'}\Vec{V}_f^{(0)}(\Vec{k}-\Vec{k}',\omega-\omega')\cdot\Vec{k}'_\perp\Vec{V}_f^{(0)}(\Vec{k}',\omega')$, and the result is
\begin{align}
\delta\Delta_I=& \lambda_1^2(2\pi)^4\int^>_{k',\omega'}\Vec{k}'_\perp\cdot\Hat{C}_f^{(0)}(\Vec{k}-\Vec{k}',\omega-\omega')\cdot\Vec{k}'_\perp\Hat{C}_f^{(0)}(\Vec{k}',\omega')\notag\\
&-\lambda_1^2(2\pi)^4\int^>_{k',\omega'}\Vec{k}'_\perp\cdot\Hat{C}_f^{(0)}(\Vec{k}',\omega')(\Vec{k}'_\perp-\Vec{k}_\perp)\cdot\Hat{C}_f^{(0)}(\Vec{k}-\Vec{k}',\omega-\omega').
\end{align}
We can estimate $\delta \Delta_I$ in the same way as we discussed the diffusion correction and always keep in mind that $\Delta=\Delta(1-e^{-k^2/\Lambda^2})\approx\Delta k^2/\Lambda^2$ when $k$ is small enough. The final result is
\begin{align}
\delta \Delta_I&\approx \Delta\Bar{\lambda}_1^2 F(s;\Lambda,\frac{A_IB}{D})\notag\\
&\approx \Delta\Bar{\lambda}_1^2(F_{31}(\Lambda)(s-1)+\frac{AB}{D}F_{33}(\Lambda)(s^3-1)+\cdots).
\end{align}
Apparently, the diffusion coefficient $D$ and the noise strength $\Delta$ are renormalized in the same way, as required by FDT. In summary, the transformation of parameters follows as
\begin{align}
D_I&=D+\delta D_I\notag\\
&\approx D[1+\Bar{\lambda}_1^2 F(s;\Lambda,\frac{A_IB}{D})]
\end{align}
\begin{align}
\Delta_I&=\Delta +\delta \Delta_I\notag\\
&\approx \Delta[1+\Bar{\lambda}_1^2 F(s;\Lambda,\frac{A_IB}{D})]
\end{align}
\begin{align}
\lambda_I=\lambda_1
\end{align}
\begin{align}
\elabel{B_result}
B_I=B
\end{align}
At last, we proceed to the second step, namely, rescale lengths, time, $\Vec{V}_\perp$ and $\delta\rho$ according to $\Vec{r}\rightarrow s\Vec{r}$, $t\rightarrow s^z t$, $\Vec{V}_\perp\rightarrow s^\chi\Vec{V}_\perp$ and $\delta\rho\rightarrow s^{\chi_\rho}\delta\rho$ respectively. The parameters transform as
\begin{align}
\elabel{D_transform}
D\rightarrow D'(s)=D_I(s) s^{z-2}
\end{align}
\begin{align}
\elabel{Delta_transform}
\Delta\rightarrow \Delta'(s)=\Delta_I(s) s^{z-2\chi-3}
\end{align}
\begin{align}
\elabel{lambda_transform}
\lambda\rightarrow \lambda_1'(s)=\lambda_1 s^{z+\chi-1}
\end{align}
\begin{align}
\elabel{B_transform}
B\rightarrow B'(s)=Bs^{z+\chi_\rho-\chi-1}
\end{align}
The recursion relations for these parameters describe how they change under the scaling transformation, which is quite obvious. To further determine scaling exponents $z$ and $\chi$, we then take an infinitesimal transformation, i.e., $s=1+l$ ($l\rightarrow0$), which leads to differential recursion relations for the parameters
\begin{align}
\elabel{flow_D}
&\frac{dD(l)}{dl}=D[z-2+F_{31}\Bar{\lambda}_1^2],\\
\elabel{flow_Delta}
& \frac{d\Delta(l)}{dl}=\Delta[z-2\chi-3+F_{31}\Bar{\lambda}_1^2],\\
\elabel{flow_lambda}
& \frac{d\lambda_1}{dl}=\lambda_1[z+\chi-1],\\
& \frac{dB(l)}{dl}=B[\chi_\rho-1+z-\chi],
 \end{align}
The scaling factors $z$, $\chi$ and $\chi_\rho$ are chosen to keep $D$, $\Delta$, $\lambda_1$ and $B$ fixed at their initial value. It could be achieved by letting the above derivatives be zero, and the resulting exponents are $z=3/2$, $\chi=-1/2$, and $\chi_\rho=-1$.

\bibliography{QVM}

\clearpage

\appendix